\documentclass{jfm}

\usepackage{graphicx}
\usepackage{natbib}
\usepackage{amsmath}
\usepackage{esint}
\usepackage{amssymb}
\usepackage{multirow}
\usepackage{xcolor}

\renewcommand{\d}[0]{\textnormal{d}}

\renewcommand{\Re}[0]{\textit{Re}}

\renewcommand{\vec}[1]{\boldsymbol{#1}}
\newcommand{\grad}{\boldsymbol{\nabla}}
\renewcommand{\div}{\boldsymbol{\nabla} \cdot}

\newcommand{\pd}[2]{\frac{\partial #1}{\partial #2}}

\newcommand{\intprod}{\mathbin{\raisebox{\depth}{\scalebox{1}[-1]{$\lnot$}}}}
\newcommand{\ignore}[1]{}

\newcommand{\mybar}[1]{\makebox[0pt]{$\phantom{#1}\overline{\phantom{#1}}$}#1}

\newcommand{\rev}[1]{#1}

\title{Unified description of turbulent entrainment}
\author[M. van Reeuwijk, J.C.\ Vassilicos and J.\ Craske]{Maarten van Reeuwijk\aff{1}, J. Christos Vassilicos\aff{2} and John Craske\aff{1}}
\affiliation{
\aff{1} Department of Civil and Environmental Engineering, Imperial College London, \\ London SW7 2AZ, UK \\
\aff{2} Univ.\ Lille, CNRS, ONERA, Arts et Metiers ParisTech, Centrale Lille, UMR 9014 - LMFL - Laboratoire de M\'ecanique des Fluides de Lille - Kamp\'e de Feriet, F-59000 Lille, France
}

\begin{document}

\maketitle

\begin{abstract}We present a mathematical description of turbulent entrainment that is applicable to free shear problems that evolve in space, time or both. Defining the global entrainment velocity \rev{$\mybar V_g$} to be the fluid motion across an isosurface of an averaged scalar, we find that for a slender  flow, $\mybar V_g=\overline u_\zeta - \overline{D}h_t/\overline{D}t$, where $\overline D/\overline D t$ is the material derivative of the average flowfield and $\overline u_\zeta$ is the average  velocity perpendicular to the flow direction across the interface located at $\zeta=h_t$.
The description is shown to reproduce well-known results for the axisymmetric jet, the planar wake and the temporal jet, and provides a clear link between the local (small-scale) and global (integral) descriptions of turbulent entrainment.
Application to unsteady jets/plumes demonstrates that, under unsteady conditions, the entrainment coefficient $\alpha$ no longer only captures entrainment of ambient fluid, but also time-dependency effects due to the loss of self-similarity.
\end{abstract}

\section{Introduction}
\label{sec:introduction}

Despite over half a century of research and several review articles
\citep[e.g.][]{Turner1986, Fernando1991, Woods2010, deRooy2013, DaSilva2014, Mellado2017}, our understanding of turbulent entrainment
(the transport of fluid from regions of relatively low to relatively
high levels of turbulence) remains fragmented. One important reason is
that turbulent entrainment is notoriously difficult to determine.
Entrainment is a process that typically occurs over much larger
timescales than turbulent timescales, and its effects are therefore
easily obfuscated by turbulent fluctuations and transient
effects. Furthermore, the quantification of turbulent entrainment
requires the determination of a turbulent and non-turbulent (or less
turbulent) region which is, by definition, arbitrary and thus subject
to uncertainty.
 
However, there are other reasons that the understanding of
turbulent entrainment remains challenging. One challenge is the
sheer number of flows in which turbulent entrainment plays a role.
Developing boundary layers can be classified based on the number
of independent variables on which their solution depends, with a
further distinction between statistically steady and unsteady
problems, as shown in table \ref{tab:flows}.  Consider a turbulent
velocity field $\vec u(x,y,z,t)$, which will generally not have
any symmetries.
By ensemble averaging this velocity field, denoted
by $\overline \cdot$, an average velocity field $\overline{ \vec u}$ is obtained which satisfies the symmetries present in the problem formulation (such
as axisymmetric or streamwise homogeneity). In table
\ref{tab:flows}, $x$ is the (slowly developing) streamwise
direction, and $z$ or $r$ is the normal direction, where $z$ would
be used for planar problems and $r$ for axisymmetric problems.
Finally, the class of steady problems with two independent variables
comprises e.g.\ planar and axisymmetric jets \citep{Hussein1994, DaSilva2002, Westerweel2005, Watanabe2014}, plumes \citep{LisEafm1982a}, wakes \citep{Cantwell1983, Obligado2016}, fountains \citep{Hunt2015}, boundary layers \citep{Head1958, Sillero2013}, mixing layers \citep{Rajaratnam1976} and inclined gravity currents \citep{Wells2010, Odier2014, Krug2015}.

In the class of unsteady problems with two independent variables are problems that develop slowly in time in one spatial dimension $z$ or $r$. 
These are problems such as penetrative convection \citep{Mellado2012, Holzner2017}, convective and stable boundary layers \citep[as relevant to the atmospheric boundary layer and the oceanic mixed layer; ][]{Kato1969, Deardorff1980, Sullivan1998, Jonker2013, Garcia2014}, stratocumulus clouds \citep{Mellado2017}, but also include temporal jets \citep{DaSilva2008,vanReeuwijk2014}, plumes \citep{Krug2017}, gravity currents \citep{vanReeuwijk2018, vanReeuwijk2019}, wakes \citep{Redford2012, Watanabe2016}, mixing layers \citep{Watanabe2018} and compressible reacting mixing layers \citep{Jahanbakhshi2018}. 
These temporal flows are not generally encountered in nature but
share many of the features of their 2D steady cousins. However, with two homogeneous spatial directions, they are ideal for exploration with direct numerical simulation.

Steady problems with three independent variables possess two
normal directions in which the flow develops `fast' but in an
anisotropic manner.  Examples are jets and plumes discharged
vertically in a crosswind \citep{Mahesh2013, DeWit2014, Woods2010,
  Devenish2010}, stratified wakes \citep{XuYphf1995a} and horizontally discharged
point releases in stratified layers.  The class of unsteady problems with three independent variables comprises all the flows mentioned in the category of 2D steady developing boundary layers, provided that
one of their boundary conditions or the environment changes in
time.  Examples include unsteady jets and plumes
\citep{ScaMjfm2006c, Craske2015, Craske2016, WooMjfm2016a} and
starting plumes \citep{TurJjfm1962a}.
The class of unsteady problems with four independent variables comprises unsteady versions in the category of 3D steady free shear flows. These comprise unsteady gravity currents from a point source, and unsteady jets/plumes in a cross-flow.
\begin{table}
\centering
\caption{Classification of free shear flows based on the number of independent variables.}
\begin{tabular}{c c c }

\hline
& & \\
Dims & Steady & Unsteady \\
\hline
1 
& \begin{minipage}{60mm} ~\\ \centering $\overline{\vec{u}}(z)$ \\ (not encountered in free shear flows) \\
\end{minipage} 
& \begin{minipage}{60mm} ~\\ \centering $\overline{\vec{u}}(t)$ \\ (not encountered in free shear flows) \\
\end{minipage}  \\ \\
2 
& \begin{minipage}{60mm} ~\\ \centering $\overline{\vec{u}}(x,z)$, $\overline{\vec{u}}(x,r)$ \\ jets, wakes, mixing layers, plumes, inclined gravity currents, $\ldots$ \newline \newline 
\end{minipage} 
& \begin{minipage}{60mm} ~\\ \centering $\overline{\vec{u}}(z,t)$, $\overline{\vec{u}}(r,t)$ \\ 
Temporal jets, wakes, mixing layers, plumes, inclined gravity currents, penetrative convection, convective boundary layer, $\ldots$ 
\end{minipage}  \\ \\
3 
& \begin{minipage}{60mm} ~\\ \centering $\overline{\vec{u}}(x,y,z)$ \\ gravity currents from a point source, stratified wakes, jets and plumes in crossflow, $\ldots$ \newline
\end{minipage} 
& \begin{minipage}{60mm} ~\\ \centering $\overline{\vec{u}}(x, z,t)$, $\overline{\vec{u}}(x, r,t)$ \\ 
Unsteady versions of those in the category of 2D steady flows, e.g.\ unsteady jets, plumes, $\ldots$
\end{minipage}  \\ \\
4 
& 
& \begin{minipage}{60mm} ~\\ \centering $\overline{\vec{u}}(x, y, z, t)$ \\ 
Unsteady versions of those in the category of 3D steady flows, e.g.\ gravity currents from a point source with variable discharge, $\ldots$
\end{minipage} 
\label{tab:flows}
\end{tabular}
\end{table}

Turbulent entrainment is generally studied either from a local or
a global perpective.  The global approach involves inferring the
entrainment velocity from the Reynolds-averaged equations
\citep[e.g.][]{Turner1986, Townsend1976} and considers entrainment from an
integral perpective.  
The local approach, as pioneered by
\citet{Corrsin1955}, considers the microscale perspective.  The
local approach starts from choosing a scalar quantity $\chi$ to
provide an implicit definition of the instantaneous
turbulent-nonturbulent interface (TNTI), where a threshold value
$\chi_0$ is used to distinguish the turbulent zone
($\chi\ge\chi_0$) from the nonturbulent zone ($\chi<\chi_0$).  The
most commonly used scalar quantity is enstrophy
\citep[e.g.][]{Bisset2002, Holzner2011, DaSilva2014,
  vanReeuwijk2014}, which is consistent with \cite{Corrsin1955},
but passive scalars \citep[e.g.][]{Sreenivasan1989,
  Westerweel2005, Burridge2017} or the turbulence kinetic energy
\citep{Philip2014, Chauhan2014} are also used to define the TNTI.
Note however that care needs to be taken when using turbulence kinetic energy to delineate the TNTI, as pressure can induce irrotational velocity fluctuations in the ambient \citep{Watanabe2018a}.

\rev{
The choice of the threshold value $\chi_0$ is always slightly arbitrary, as the flow transitions smoothly from turbulent to non-turbulent.
In addition any measurement and simulation data are subject to uncertainty, and background levels of $\chi$ may be nonzero (e.g. enstrophy levels in a turbulent ambient). One therefore typically chooses a small and finite nonzero threshold $\chi_0$ to define the TNTI.
Since the interface between turbulent and non-turbulent fluid is generally very sharp, there is a range of thresholds $\chi_0$ that can be chosen for which the entrainment statistics are insensitive to the choice of $\chi_0$ \citep{DaSilva2014}.
}

\begin{figure}
\centering
\includegraphics[width=11cm]{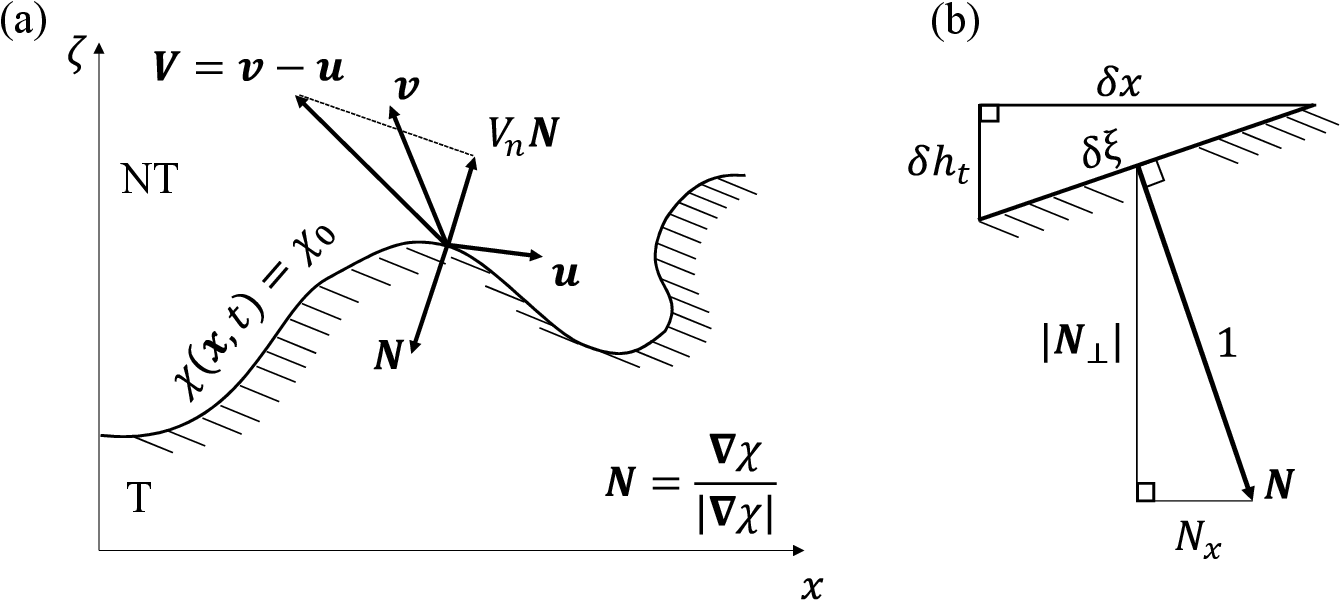}
\caption{Interface properties of the turbulent-nonturbulent interface (TNTI) separating turbulent (T) from non-turbulent (NT) fluid. (a) Definition sketch. (b) Geometric properties.}
\label{fig:local}
\end{figure}

The velocity $\vec{v}$ associated with any trajectory on an
isosurface of $\chi$ satisfies 
\begin{equation}
\label{eq:chi_iso}
  \frac{\d \chi}{\d t} = \frac{\partial \chi}{\partial t} + \vec{v} \cdot \grad\chi = 0.
\end{equation}
By introducing a relative velocity $\vec V  = \vec v - \vec u$, which is the difference between the isosurface velocity $\vec v$ and the fluid velocity $\vec u$, \eqref{eq:chi_iso} can be rewritten as \citep{Dopazo2007, Holzner2011}
\begin{equation}
\label{eq:Vn}
V_n = - \frac{1}{| \grad \chi |} \frac{D \chi}{D t},
\end{equation}
where $D/Dt=\partial/\partial t + \vec u \cdot \grad$ is the
material derivative, $V_n = \vec V \cdot \vec N$ is the normal
component of the relative velocity and
$\vec N=\grad \chi / | \grad \chi|$ is the (3D) normal vector
pointing into the turbulent region (Figure \ref{fig:local}(a)).
Note that the other two components of $\vec V$ tangential to the
isosurface are not specified by this definition.  For an
entraining flow, $V_n<0$, which is a consequence of defining a
normal $\vec N$ that points \emph{into} $\Omega$.  The inward
pointing normal also has consequences for the Gauss divergence
theorem.
By substituting the governing equation for $\chi$ (usually enstrophy) into \eqref{eq:Vn} and averaging, local aspects of turbulent entrainment can be explored \citep[e.g.][]{Holzner2011, DaSilva2014, vanReeuwijk2014, Krug2015, Jahanbakhshi2018}.

The global entrainment velocity $\mybar V_g$ is not as uniquely defined as the local entrainment velocity \eqref{eq:Vn} \citep{Turner1986, Hunt1983}.
For spatially developing flows, the entrainment velocity is usually associated with a flow into the turbulent region. For temporal problems however, it is defined via $\d h / \d t$, the growth of the layer \rev{in time}, where $h$ is some characteristic layer thickness.
However, for spatially developing flows in which the environment is non-quiescent or is stratified, there is another mode of entrainment which is associated with the entrainment across the boundary.  
These different forms of entrainment were discussed in \cite{Hunt1983, Turner1986} and cause confusion between disciplines, as they describe related processes that are not necessarily equivalent. In this paper we derive an unambiguous definition of the global entrainment velocity that can be used for spatial, temporal and spatio-temporal (unsteady) entrainment problems.

%

The aim of this paper is to derive an integral description of free shear flows capable of representing both the local and global viewpoints of turbulent entrainment.
An equivalent definition to \eqref{eq:Vn} is presented for the global entrainment velocity.
The framework provides a unified description of entrainment in temporal problems  \citep[2D unsteady;][]{DaSilva2008, vanReeuwijk2014, Krug2017}, in which the TNTI moves but there is no net flow into the turbulent layer ($V_n$ produced by $\vec v$), spatial problems \citep[2D steady;][]{Rajaratnam1976, Turner1986, Philip2014}, in which the TNTI is statistically steady but there is a net flow into the turbulent layer ($V_n$ produced by $\vec u$), and unsteady free shear layers \citep[3D unsteady;][]{Craske2016}.

This paper is organised as follows. \S \ref{sec:cont} introduces the integral operator identities and the averaged plane-integrated Navier-Stokes equations which describe the integral spatio-temporal dynamics of free shear flows.
In \S \ref{sec:global}, an average isosurface of $\overline \chi$ is applied to the equations which results in an expression for the global entrainment velocity $\mybar V_g$ in terms of an explicit function $h_t$ describing the TNTI.
The implications of this new definition of $\mybar V_g$ are discussed in \S \ref{sec:implications}.
Four canonical free shear flows (an axisymmetric jet, a planar wake, a temporal jet and an unsteady jet/plume) are studied in \S \ref{sec:cases} to show that the current definition of $\mybar V_g$ is fully consistent with previous results.
Application of the new description to unsteady jets and plumes reveals the relation between the entrainment coefficient $\alpha$ and actual entrainment across the interface.
Concluding remarks are made in \S \ref{sec:conclusions}.

\section{Local averaged integral equations}
\label{sec:cont}

In this section we present integral volume
(i.e. continuity) and momentum conservation equations. The focus of
this work is on turbulent free shear flows which develop slowly (in a
statistical sense) in time $t$, in the spatial direction $x$, or both.
The incompressible Navier-Stokes equations are given by
\begin{align}
\label{eq:cont}
\div \vec u &=0, \\
\label{eq:mom}
\frac{D \vec u}{D t}  &= - \frac{1}{\rho_0} \grad p + \div \vec  \tau + \vec f,
\end{align}
where $\vec \tau$ denotes the viscous stress tensor and $\vec f$ is a body force. 
These equations will be integrated over the turbulent region in the $y-z$ plane which is denoted $\Omega$.
The following identities for the integrals of the gradient, divergence and material derivative operators can be derived:
\begin{align}
  \label{eq:gradint}
  \int_{\Omega} \grad \phi \d A 
  & = \frac{\partial}{\partial x} \int_{\Omega} \phi \d A \vec e_x
    - \oint_{\partial \Omega} \phi \frac{\vec N}{| \vec N_\perp|} \d \ell, \\
    \label{eq:divint}
\int_{\Omega} \div \vec F \d A &= \frac{\partial}{\partial x}\int_{\Omega} F_x \d A- \oint_{\partial \Omega} \vec{F} \cdot \frac{\vec N}{| \vec N_\perp|} \d \ell, \\
\label{eq:matderint}
\int_{\Omega} \frac{D \phi}{D t} \d A &= 
\frac{\partial }{\partial t}\int_{\Omega} \phi ~ \d A  +    \frac{\partial}{\partial x} \int_{\Omega} u_x \phi ~ \d A 
- \int_{\Omega} \phi \div \vec u  \d A
+ \oint_{\partial \Omega}  \frac{V_n}{| \vec N_\perp|} \phi ~ \d \ell,
\end{align}
where $\phi$ is an arbitrary scalar or vector component field and
$\vec F$ is an arbitrary vector field. Vectors with a perp ($\perp$)
subscript denote the components perpendicular to the $x$-direction,
i.e.  $\vec F =[F_x, \vec F_\perp]^T$, and $\grad_\perp
=[\partial/\partial y, \partial / \partial z]^T$. The unit vector
$\vec N=\grad \chi / | \grad \chi|$ is normal to the 3D surface $\chi
= \chi_0$ which demarcates between turbulent and non-turbulent
regions. The normal vector can be written as $\vec N = [N_x, \vec
  N_\perp]^T$, so that $|\vec N_\perp|$ is the magnitude of the 3D
normal in the $y-z$ plane [see figure \ref{fig:local}(b)].  Finally,
$\vec{e}_x$ is the unit vector in the $x$-direction and $u_x$ is the
component of the fluid velocity field $\vec u$ in that same direction.
An easily accessible yet rigorous proof of these three identities is
given in Appendix \ref{sec:intop}.  A more general derivation using
differential geometry, which highlights the role of Stokes' theorem
and the Leibniz integral rule, is given in Appendix \ref{sec:dg}.

\rev{
Since the flow is turbulent, the integration domain $\Omega(y,z;x)$ can consist of several disconnected blobs of turbulent fluid, i.e.\ $\Omega = \Omega_1 \cup \Omega_2 \cup \ldots $. This implies that the domain boundary $\partial \Omega$ contains multiple closed trajectories $\partial \Omega_1, \partial \Omega_2, \ldots $ which are summed up with the line integral, That is, $\oint_{\partial \Omega}\cdot \d \ell =\oint_{\partial \Omega_1}\cdot \d \ell +\oint_{\partial \Omega_2}\cdot \d \ell +\ldots$ if the domain contains multiple disconnected blobs.
}

Noting that $\div \vec u = D\phi /D t=0$ for $\phi = 1$ and using \eqref{eq:matderint} implies that the instantaneous integral continuity equation is given by
\begin{equation}
  \label{eq:contint}
 \frac{\partial}{\partial t} \int_{\Omega}  ~ \d A  
+ \frac{\partial}{\partial x} \int_{\Omega} u_x ~ \d A 
= -\oint_{\partial \Omega}  \frac{V_n}{| \vec N_\perp|} ~ \d \ell .
\end{equation}
Note that if the relative isosurface velocity $\vec V$ is everywhere tangential
to the interface then $V_n=0$, in which case \eqref{eq:contint} describes a streamtube. 
Entrainment allows exchange across the isosurface $\chi=\chi_0$.
For an entraining flow, $V_n<0$ (due to the inward pointing normal).

The line integral represents the net entrainment into the turbulent
region. The factor $|\vec N_\perp|$ accounts for the
projection of the 3D quantity $V_n$ onto the $y-z$ plane. This is
better seen by using $V_n = \vec V \cdot \vec N$ to write the
entrainment term as
\begin{equation}
\label{eq:Vndecomp}
\oint_{\partial \Omega}  \frac{V_n}{| \vec N_\perp|} ~ \d \ell = \oint_{\partial \Omega}  \vec V_\perp \cdot \vec n ~ \d \ell
+ \oint_{\partial \Omega}  V_x\frac{N_x}{| \vec N_\perp|} ~ \d \ell, 
\end{equation}
where $\vec n = \grad_\perp \chi/|\grad_\perp \chi |$ is the
normal in the $y-z$ plane, and note that this quantity is related
to the 3D normal via $\vec n = \vec N_\perp / |\vec N_\perp|$.
The first term on the right-hand side of \eqref{eq:Vndecomp} is simply
the entrainment flux arising from the in-plane relative velocity
components $\vec V_\perp$. As described in appendices
\ref{sec:intop} and \ref{sec:dg}, the second term on the right-hand
side of \eqref{eq:Vndecomp} arises from the commutation of integration
and differiation with respect to $x$ that is required to formulate
\eqref{eq:contint}. It represents the net transport of the
streamwise relative velocity component $V_x$ into the turbulent
region across the interface whose local slope is
$N_x/|\vec N_\perp|$ (see Figure \ref{fig:local}).
\rev{We note that $|\vec N_\perp|$ can be zero locally when $\vec N$ is aligned with the $x$-direction , which would render the integrand infinite. 
This is an inescapable consequence of determining entrainment as a function of $x$. Any subsequent integration over $x$ will remain finite however, since $V_n/ | \vec N_\perp| \d \ell \d x =V_n \d S$ where $\d S$ is the local surface area of the surface in 3D.}  

Integration over the region $\Omega$ of \eqref{eq:mom} and use of identities \eqref{eq:gradint}-\eqref{eq:matderint} results in
\begin{equation}
  \label{eq:momint}
\begin{split}
\frac{\partial}{\partial t} \int_{\Omega} \vec{u} ~ \d A  &+ \frac{\partial}{\partial x} \int_{\Omega} \left(u_x \vec{u}  +\frac{p}{\rho_0} \vec e_x\right) ~ \d A 
= \\ &- \oint_{\partial \Omega}  \frac{1}{| \vec N_\perp|} \left( V_{n} \vec u - \frac{p}{\rho_0} \vec N\right) ~ \d \ell 
+ \int_{\Omega} \vec f \d A.
\end{split}
\end{equation}
Here, the shear-stress contributions have been neglected as is conventional for high Reynolds free shear flows.
Equations \eqref{eq:contint} and \eqref{eq:momint} are instantaneous.


Performing ensemble averaging, denoted by the overbar $\overline
\cdot$, on the instantaneous integrated continuity equation
\eqref{eq:contint} and the streamwise $x-$component of the integrated
momentum equation \eqref{eq:momint} yields
\begin{gather}
  \label{eq:contintav}
 \frac{\partial}{\partial t} {\overline{\int_{\Omega}   ~ \d A}}
+ \frac{\partial}{\partial x} \overline{\int_{\Omega} u_x  ~ \d A }
 = -\overline{\oint_{\partial \Omega}  \frac{V_{n}}{|\vec N_\perp|} ~ \d \ell}, \\
  \label{eq:momintav}
\begin{split}
\frac{\partial}{\partial t} \overline{\int_{\Omega} u_x ~ \d A}  &+ \frac{\partial}{\partial x} \overline{\int_{\Omega} \left(u_x^2 + \frac{p}{\rho_0}\right) ~ \d A} =\\
& - \overline{\oint_{\partial \Omega}  \frac{1}{| \vec N_\perp|} \left( V_{n} u_x - \frac{p}{\rho_0} N_x \right) ~ \d \ell}  + \overline{\int_{\Omega} f_x \d A}.
\end{split}
\end{gather}
In the integral continuity equation \eqref{eq:contintav}, $\overline{\int_{\Omega} \d A}$ represents the average instantaneous cross-sectional area of the turbulent region at location $x$.
It is not possible to commute the integral with the ensemble averaging because the integration regions $\Omega$ and $\partial \Omega$ vary in time and per ensemble instance. 

\section{Global averaged integral equations}
\label{sec:global}
\begin{figure}
\centering
\includegraphics[width=6cm]{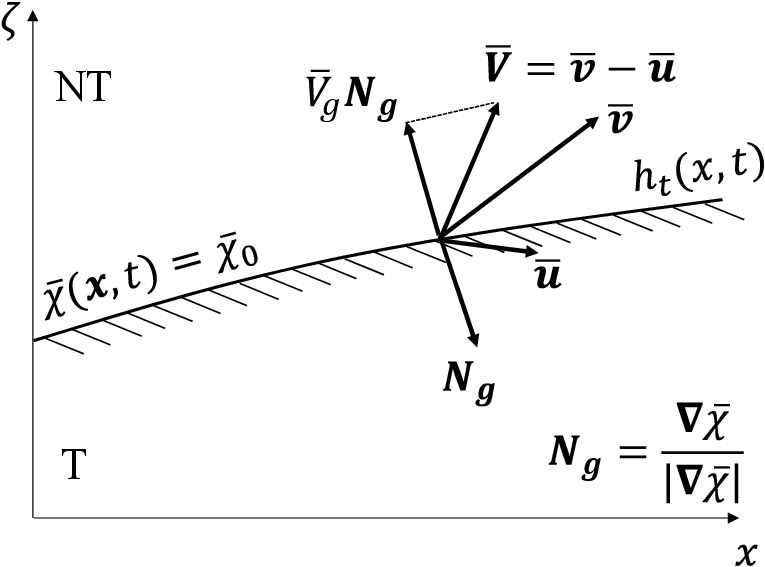}
\caption{Definition sketch of the global (average) perspective on the  turbulent-nonturbulent interface (TNTI).}
\label{fig:global}
\end{figure}

Equations \eqref{eq:contintav}, \eqref{eq:momintav} ultimately
link the integral behaviour of the free shear flow to the
small-scale dynamics at the TNTI when $\chi$ is an instantaneous
quantity.  
The global, integral dynamics can be obtained by using
an \emph{average} quantity $\overline \chi$, with associated threshold $\overline \chi_0$, to identify the
interface (Figure \ref{fig:global}).  
By considering an average quantity, the TNTI will not
be contorted but will be smooth and satisfy the symmetries
corresponding to homogeneity in the problem under consideration.

\subsection{Non-slender flows}
We define the global entrainment velocity $\mybar V_g$ to be the net transport across an averaged scalar $\overline{\chi}$. This implies that the 3D normal is defined as
$\vec N_g = \grad \overline \chi / | \grad \overline \chi |$, and therefore that
\begin{equation}
\label{eq:Vnavdef}
V_g = ({\vec{v}}-{\vec{u}}) \cdot \vec{N}_g
= \frac{{\vec v} \cdot \grad \overline \chi - {\vec u} \cdot \grad \overline \chi}{|\grad \overline \chi|} 
= - \frac{1}{|\grad \overline \chi|} \frac{ D \overline \chi}{ D t},
\end{equation}
where use was made of \eqref{eq:chi_iso} for $\chi = \overline\chi$.
The equation above is the \emph{instantaneous} global entrainment velocity across an isosurface based on a Reynolds-averaged quantity. The mean entrainment velocity can be determined by applying Reynolds-averaging to \eqref{eq:Vnavdef}, with result
\begin{equation}
\label{eq:Vnavavdef}
\mybar V_g =  \overline{\vec V \cdot \vec N_g} 
=   \overline{\vec V} \cdot \vec N_g = - \frac{1}{|\grad \overline \chi|} \frac{\overline D \overline \chi}{\overline D t},
\end{equation}
where $\overline D / \overline D t = \partial/\partial t + \overline{\vec u} \cdot \grad$.

An advantage of considering averaged quantities is that it is possible to represent the isosurface $\overline \chi=\overline \chi_0$ (which implicitly defines the TNTI) \emph{explicitly} in terms of a single-valued function $h_t$ in a coordinate system appropriately representing the symmetries of the free-shear problem under consideration.
Restricting attention to planar or axisymmetric problems, a scalar level-set function $L(x,\zeta ,t) = h_t(x, t) - \zeta$ can be constructed such that $L=0$ represents the average interface position where $\zeta$ is the direction normal to the $x$-direction. 
Setting $\overline{\chi} = L$ implies that \eqref{eq:Vnavdef} can equivalently be expressed as 
\begin{equation}
\label{eq:Vnavh}
\mybar V_g =  -\frac{1}{|\grad L |} \left( \frac{\overline D h_t}{\overline D t} - \frac{\overline D \zeta}{\overline D t}\right) 
= \frac{1}{| \grad L |} \left( \overline u_\zeta - \frac{\overline D h_t}{\overline D t} \right),
\end{equation}
where $|\grad L| = \sqrt{1 + (\partial h_t / \partial x)^2}$.

Because the interface is based on the average quantity $\overline{\chi}$, the averaging and integral operators \emph do commute, which implies that  \eqref{eq:contintav}, \eqref{eq:momintav} simplify to
\begin{gather}
  \label{eq:contintavG}
 \frac{\partial}{\partial t} {\int_{\overline{\Omega}} \d A}
+ \frac{\partial}{\partial x} \int_{\overline{\Omega}} \overline{u}_x ~ \d A 
 = -\oint_{\partial \overline{\Omega}} \frac{\mybar{V}_{g}}{|\vec N_{g\perp}|} ~ \d \ell, \\
  \label{eq:momintavG}
\begin{split}
\frac{\partial}{\partial t} \int_{\overline{\Omega}} \overline{u}_x ~ \d A  & + \frac{\partial}{\partial x} \int_{\overline{\Omega}} \left(\overline{u_x^2} + \frac{\overline{p}}{\rho_0} \right) ~ \d A =\\
& - \oint_{\partial \overline{\Omega}}  \frac{1}{| \vec N_{g\perp}|} \left( \overline{V_{g} u_x} - \frac{\overline{p}}{\rho_0} N_{gx} \right) ~ \d \ell  + \int_{\overline{\Omega}} \overline{f}_x \d A.
\end{split}
\end{gather}
Here, we have denoted the integration domain and boundary with $\overline{\Omega}$ and $\partial \overline{\Omega}$, respectively, to distinguish from the local viewpoint.
The integrals can be made definite once a specific coordinate system is selected.
Note that $\overline{u_x^2}=\overline{u}_x^2 + \overline{u'_x u'_x}$.

\subsection{Slender flows}
Many free shear flows have the additional property of being
\emph{slender}, i.e.\ they develop much more slowly in the
streamwise $x$-direction than in the normal direction, which
implies that, apart from all quantities changing slowly in the
$x$-direction,
$\partial \overline \chi/\partial x \ll | \grad_\perp
\overline{\chi} |$.  Under the assumption of slenderness,
\eqref{eq:Vnavdef}, \eqref{eq:Vnavh} become
\begin{equation}
\label{eq:Vnav}
\mybar{V}_g  
=   - \frac{1}{|\grad_\perp \overline \chi|} \frac{\overline D \overline \chi}{\overline D t} 
=  \overline u_\zeta - \frac{\overline D h_t}{\overline D t}.
\end{equation}
The assumption of slenderness also implies that $|N_{gx}| \ll |\vec N_{g\perp} |$, which furthermore implies that $|\vec N_{g\perp} | \approx 1$ and $\vec N_{g\perp} \approx \vec n_g=\grad_\perp \overline \chi/|\grad_\perp \overline \chi|$.
Thus, under this assumption the equations (\ref{eq:contintavG},\ref{eq:momintavG}) further simplify to  
\begin{gather}
  \label{eq:contintavGS}
 \frac{\partial}{\partial t} {\int_{\overline{\Omega}}  ~ \d A}
+ \frac{\partial}{\partial x} \int_{\overline{\Omega}} \overline{u}_x   ~ \d A 
 = -\oint_{\partial \overline{\Omega}}  \mybar{V}_{g} ~ \d \ell, \\
  \label{eq:momintavGS}
\begin{split}
\frac{\partial}{\partial t} \int_{\overline{\Omega}} \overline{u}_x  ~ \d A  &+ \frac{\partial}{\partial x} \int_{\overline{\Omega}} \left( \overline{u_x^2}+\frac{\overline p }{\rho_0} \right) ~ \d A \\
&= 
- \oint_{\partial \overline{\Omega}}  \left( \overline{V_{g} u_x} - \frac{\overline{p}}{\rho_0} N_{gx} \right)  ~ \d \ell  
+ \int_{\overline{\Omega}} \overline{f}_x \d A.    
\end{split}
\end{gather}

\section{Implications}
\label{sec:implications}
\subsection{Reconciliation of entrainment definitions}
Equation \eqref{eq:Vnav} brings perspective to the different definitions of the global entrainment velocity that have previously been used. Taking the example of an axisymmetric jet, it follows that $\zeta=r$ and thus that \eqref{eq:Vnav}  is given by
\begin{equation}
    \mybar{V}_g(x,t) = \overline u_r(x, h_t, t)  -\left(\frac{\partial h_t}{\partial t}(x, t) + \overline u_x(x,h_t, t) \frac{\partial h_t}{\partial x}(x, t)\right).
    \label{eq:Vnav_expanded}
\end{equation}
\cite{Hunt1983, Turner1986}, discuss the different definitions that were used in the thirty years prior, and distinguished between the  entrainment rate $E$, a boundary entrainment rate $E_b$ and a net entrainment rate $E^*$.
Loosely speaking, one can relate $\mybar V_g$ with the net entrainment rate $E^*$, $\overline u_r$ with the entrainment rate $E$ and $\overline D h_t / \overline D t$ with the boundary entrainment rate $E_b$, such that $E^* = E - E_b$.
However, note that only steady problems were considered in these discussions.
The definitions of $E^*$, $E$ and $E_b$ will be detailed below, as well as the similarities and differences in the concepts.

In \cite{Hunt1983, Turner1986}, $E$ and $E_b$ were determined from the top-hat "cartoon" of a jet, i.e.\ the jet has a width $h$ with a uniform velocity $u_m$ inside and zero velocity outside it \rev{(this viewpoint has been used extensively for integral descriptions of free shear flows and should not be applied locally, e.g.\ to quantify Reynolds stresses which would be zero). }
Here, the top-hat width $h$ and velocity $u_m$ are defined as $h=Q/M^{1/2}$ and $u_m=M/Q$, where $Q = 2 \int_0^{h_t} \overline u_x r \d r$ is the volume flux and $M = 2 \int_0^{h_t} \overline u_x^2 r \d r$ is the momentum flux per unit radian. For a steady jet, $\d h / \d x = 2 \alpha$, where $\alpha$ is the entrainment rate \citep{Turner1986}.
In the cartoon view, $E=-\alpha u_m$ is the flow perpendicular through the jet boundary, and $E_b = u_m \d h / \d x$ is the outward velocity of an observer moving along the interface.
Using the spreading rate of the jet, we can write this as $E_b = 2 \alpha u_m = 2 E$, such that the net entrainment rate $E^*=-3E$.

However, the classical arguments suggest that there is a choice in the  entrainment definition when \eqref{eq:Vnav_expanded} clarifies there is  not. Indeed,  if we restrict ourselves to a steady flow and choose a conventional threshold strategy by using $\overline \chi = \overline u(x,r) / u_m(x)$ \citep[e.g.][]{vanReeuwijk2016}, then we immediately obtain that 
\begin{equation}
    \mybar{V}_g(x,t) = \overline u_r(x, h_t)  - \underbrace{\overline \chi_0 u_m  \frac{\d h_t}{\d x}}_{\approx 0},
\end{equation}
where the boundary contribution can be assumed zero because $\overline \chi_0 \ll 1$. This then immediately implies that $E_b\approx 0$, and that the only correct expression for the entrainment rate for the steady axisymmetric jet is that \rev{$E^* = E$}. 

\subsection{Relation between $h$ and $h_t$}

Integral models do not typically make explicit reference to a scalar interface $h_t$. Instead, it is conventional to define a characteristic width of the flow $h$ from either integral flow properties, resulting in a (top-hat) width $h=Q/M^{1/2}$ (as used in the previous section), or via the specific features of a given velocity profile, such the half-width $h_{1/2}$, defined as $\overline u_{x}(x,h_{1/2})=\overline{u}_{x}(x,0)/2$ where $\overline u_{x}(x,0)$ and is the centreline velocity.

For self-similar flows, $h$ and $h_{1/2}$ and $h_t$ are trivially related by a proportionality coefficient which translates through to the value of the entrainment coefficient. Therefore, they play an
influential, albeit superficial, role in studies of
entrainment. A further complication arises in flows that
are not self-similar, which makes it impossible to relate $h$ with
$h_{t}$ using a constant of proportionality (see \S
\ref{sec:unsteady})

Note that $h_t$ is present in the definitions of the the volume flux $Q$ and momentum flux $M$. This is important from a practical perspective as the flow in the ambient, even if only considering the induced
irrotational flow due to entrainment, will not necessarily produce a finite volume flux from integrals to infinity
\citep{KotNjfm1978a} and will contaminate the results.

\subsection{Entrainment interacting with turbulence}
Although turbulent entrainment is usually associated with a net velocity relative to an interface to the interior turbulence, it is important to acknowledge that in turbulent ambients there are also turbulent-turbulent exchanges across the interface. These are not present in the continuity equation, which does not contain products, but these terms do feature in the momentum and scalar equations.
For example, in \eqref{eq:momintavGS}, the term $\overline{V_g u_x}=\mybar{V}_g \overline{u}_x + \overline{V_g' u_x'}$ contains two contributions, one associated with the mean flow and one with turbulent exchanges across the interface. 
These can be expected to be important in environments in which there are substantial fluctuations in the ambient, such as jets in a turbulent environment \citep{Ching1995, Gaskin2004, Kankanwadi2019}, turbulent fountains \citep{Hunt2015} or clouds \citep{deRooy2013}. Importantly, the turbulent transport $\overline{V_g' u_x'}$ may require different  parameterisation than the mean transport $\mybar{V}_g \overline{u}_x$.

\subsection{Connection between the local and global viewpoints}
\label{sec:localglobal}
The integral representations of the local and global  continuity equations, \eqref{eq:contintav} and \eqref{eq:contintavG} respectively, can be used to establish the relation between the local entrainment velocity and the global entrainment velocity.
\rev{If the threshold $\chi_0$ encompasses all of the turbulence and $\overline{\chi}_0$ accounts for the mean area of the turbulent region, the left-hand sides of \eqref{eq:contintav} and \eqref{eq:contintavG} can be assumed to be  approximately equal.}
This results in
\begin{equation}
    \label{eq:localglobal}
    \overline{\oint_{\partial \Omega}\frac{V_{n}}{|\vec N_\perp|} ~ \d \ell} = 
    \oint_{\partial \overline{\Omega}}  \frac{\mybar{V}_{g}}{|\vec N_{g\perp}|} ~ \d \ell. 
\end{equation}
Consistent with \cite{Zhou2017}, we introduce the average instantaneous interface and average interface lengths as 
$\mathcal L = \overline{\oint_{\partial \Omega} \d \ell}$ and 
$\mathcal L_g = \oint_{\partial \overline{\Omega}} \d \ell$, respectively. 
Note that $\mathcal L_g$ can be determined straightforwardly from the problem geometry and $h_t$ (see also \S \ref{sec:cases}). The average instantaneous interface length ${\mathcal L}$ is expected to scale in a fractal manner \citep{Sreenivasan1989}, implying that $\mathcal L \gg \mathcal L_g$ for $Re \gg 1$.
Equation \eqref{eq:localglobal} can be recast as 
 \citep{vanReeuwijk2014, Zhou2017}
\begin{equation}
\label{eq:globallocal}
\frac{\langle \mybar V_g \rangle}{\langle V_n \rangle} = \frac{\mathcal L}{\mathcal L_g},
\end{equation} 
where $\langle V_n\rangle = \mathcal L^{-1} \overline{\oint_{\partial \Omega}
  V_n / |\vec N_\perp|\d \ell}   $ and $\langle \overline
V_g\rangle = {\mathcal L_g}^{-1} \oint_{\partial \overline{\Omega}} \mybar{V}_g /
|\vec N_{g\perp}| \d \ell $ are the effective
local and global entrainment velocity, respectively.  The fractal
arguments for $\mathcal L$ can imply that the local entrainment
velocity $\langle V_n \rangle$ is of the order of the Kolmogorov
velocity \citep{Corrsin1955, vanReeuwijk2014, Silva2018} for a
specific value of the fractal dimension of the interface, but
\cite{Zhou2017} also argued for the possibility of a different
scaling, independent of the value of the fractal dimension, in the
presence of non-equilibrium turbulence. It must be noted, however,
that the definition of local entrainment velocity used by
\cite{Zhou2017} actually relates to a pseudo-velocity (see Appendix A
and section B.2 of Appendix B). Even so, their local entrainment
velocity does scale with the Kolmogorov velocity in the presence of
classical equilibrium turbulence.
The connection between local and global entrainment was shown to hold reasonably well for an experimental study of a developing boundary layer \citep{Chauhan2014}.

\rev{
It is unlikely that \eqref{eq:localglobal} will hold in an exact manner as $\overline \chi_0 \rightarrow 0$, since global entrainment implicitly accounts for fluid entering or leaving non-turbulent regions where $\chi_{0}<\overline{\chi}<\overline{\chi}_{0}$. Indeed, \cite{Burridge2017} found that about five percent of the volume flux of a plume occurs outside of the turbulent region. For flows which are spatially and temporally evolving, deviations will likely be higher. Nevertheless, \eqref{eq:localglobal} is useful from a conceptual and practical point of view, because global entrainment is relatively straightforward to compute.}

\section{Application to four canonical flows}
\label{sec:cases}
In this section the integral description will be applied to four canonical free shear flows  namely the axisymmetric jet, the planar wake, the temporal jet and the unsteady jet/plume.
The first three cases serve to demonstrate that the framework reproduces the appropriate entrainment velocities and well-known equations and results. The fourth case, the unsteady jet/plume, will provide new insight into the interpretation of the entrainment coefficient $\alpha$.
As turbulent free shear flows are characterised by a high Reynolds number $\Re$ and a slow development in the $x$ (or $t$) direction, viscous stresses and pressure are neglected.
Furthermore, consistent with general practice on thresholding, all quantities containing a prefactor $\overline \chi_0$ will neglected.

\subsection{Axisymmetric jet}

The axisymmetric jet is homogeneous in the azimuthal direction $\theta$ and time $t$. The streamwise velocity $\overline{u}_x$ is used to define the turbulent region for the global entrainment as  $\overline \chi = \overline u_x/u_m(x)$, where $u_m(x)$ is the characteristic velocity inside the jet. Applying the symmetries to \eqref{eq:Vnav} and setting $\zeta = r$, we have
\begin{equation}
\label{eq:VnavJ}
\mybar V_g  
=  \overline u_r(x, h_t).
\end{equation}
Thus, \eqref{eq:contintavGS}, \eqref{eq:momintavGS} are given by, using that $\d A = 2 \pi r \d r$ and $\mathcal L_g = 2 \pi h_t$:
\begin{gather}
  \label{eq:contintavGJ}
2 \frac{\d}{\d x} \int_0^{h_t} \overline{u}_x r \d r
 = -2 h_t \mybar{V}_{g} = - 2 h_t \overline u_r, \\
  \label{eq:momintavGJ}
2 \frac{\d}{\d x} \int_0^{h_t} \overline{u_x^2} ~ r \d r 
= 0,
\end{gather}
which is consistent with straightforward integration of the Reynolds-averaged boundary layer equations \citep[e.g.][]{Rajaratnam1976}, thereby confirming the appropriateness of the description. 

\subsection{Planar wake}

The planar wake is an interesting case, since it features a nonzero ambient flow of amplitude $U_\infty$. This problem is statistically homogeneous in $y$ and $t$. 
Applying the symmetries to \eqref{eq:Vnav}, setting $\zeta=z$ and using $\overline \chi = 1-\overline u_x/U_\infty$ as the quantity for thresholding, we obtain
\begin{equation}
\label{eq:VnavW}
\mybar V_g  
=  \overline u_z - U_\infty \frac{\d h_t}{\d x}.
\end{equation}
Using that $\d A = L_y \d z$ and $\mathcal L_g = 2 L_y$ (since the interface is present on both sides of the $z=0$ plane), \eqref{eq:contintavGS} and \eqref{eq:momintavGS} are given by
\begin{gather}
  \label{eq:contintavGW}
 \frac{\d}{\d x} \int_{-h_t}^{h_t} \overline{u}_x \d z
 = -2 \mybar{V}_{g} = 2 \left( U_\infty \frac{\d h_t}{\d x} - \overline u_z \right), \\
  \label{eq:momintavGW}
  \begin{split}
\frac{\d}{\d x} \int_{-h_t}^{h_t} \overline{u_x^2} ~ \d z 
&= 2  U_\infty \left( U_\infty \frac{\d h_t}{\d x} - \overline u_z \right).     \\  
  \end{split}
\end{gather}
By substituting \eqref{eq:contintavGW} into \eqref{eq:momintavGW}, assuming that $\overline{u'_x u'_x} \ll \overline{u}_x^2$ and rearranging it follows that the mean momentum deficit $\int_{-h_t}^{h_t} \overline u_x ( U_\infty - \overline u_x) \d z$ is conserved as expected \citep[e.g.][]{Pope2000}.


\subsection{Temporal jet}

The capability to directly quantify the entrainment velocity $\mybar V_g$ in temporal free shear flows (e.g. atmospheric boundary layers) is one of the useful results of the integral description put forward here.
The distinguishing aspect of these flows is that they tend to be homogeneous in $x$ and $y$.
For a temporal jet, in order to obtain an expression for the global entrainment velocity we can define a threshold $\overline{\chi} = \overline u_x / u_m(t)$, where $u_m(t)$ is the characteristic value inside the jet  \citep{vanReeuwijk2014}. Applying the symmetries of this flow to \eqref{eq:Vnav} then results in
\begin{equation}
\label{eq:VnavT}
\mybar V_g  
= - \frac{\d h_t}{\d t}.
\end{equation}
Using that $\d A = L_y \d z$ and $\mathcal L_g = 2 L_y$, \eqref{eq:contintavGS} and \eqref{eq:momintavGS} simplify to
\begin{gather}
  \label{eq:contintavGT}
 2 L_y \frac{\d h_t}{\d t} 
 = -2 L_y \mybar V_{g} = 2 L_y \frac{\d h_t}{\d t}, \\
  \label{eq:momintavGT}
\begin{split}
L_y\frac{\d }{\d t} \int_{-h_t}^{h_t} \overline{u}_x ~ \d A  &=  0.
\end{split}
\end{gather}
The first equation simply confirms that $\mybar V_g$ has been defined appropriately, whilst the second equation demonstrates the conservation of volume flux for this flow \citep{vanReeuwijk2014}.

The equivalence between the integrals of local and global entrainment \eqref{eq:globallocal}, was studied in \cite{vanReeuwijk2014}.
It was shown that for the temporal jet, the entrainment at the global and local level are indeed identical over several decades of variation in $\chi_0$ (enstrophy in this case), provided it was small enough.
However, for the relatively low Reynolds number under consideration it was shown to be important to take into account the change in the interface location upon changing the threshold value $\chi_0$ if one were to determine the entrainment coefficient $\alpha$ from the local entrainment velocity. 
The consistency between the integral global and local entrainment flux \eqref{eq:globallocal} was shown also for the case of penetrative convection \citep{Holzner2017} and an inclined temporal gravity current \citep{vanReeuwijk2018}.

\subsection{Unsteady axisymmetric jets and plumes}
\label{sec:unsteady}
In this section we apply the description to unsteady axisymmetric jets and plumes, which will provide new insight in the extent to which the entrainment coefficient $\alpha$ is linked to actual entrainment across the jet/plume boundary.
Axisymmetric statistically unsteady jets and plumes retain a dependence on three independent variables: the streamwise direction $x$, the lateral or normal direction $r$, and time $t$.
In the case of unsteady jets and plumes, it cannot be assumed that the flow remains slender, since there can be substantial variation of all the quantities of interest over short distances. As for the axisymmetric jet, the streamwise velocity $\overline{u}_x$ is used to define the turbulent region for the global entrainment as  $\overline \chi = \overline u_x/u_m(x,t)$, where $u_m(x)$ is the characteristic velocity inside the jet. Applying the symmetries to \eqref{eq:contintavG}, setting $\zeta = r$ and $\d A = 2 \pi r \d r$ we obtain 
\begin{equation}
\pd{h_{t}^{2}}{t}
+ \pd{Q}{x}
 = -\frac{1}{\pi}\oint_{\partial \overline{\Omega}}  \frac{\mybar{V}_{g}}{| \vec N_{g\perp} |} ~ \d \ell, \\
\label{eq:uplume_l}
\end{equation}
where $Q=2\int_0^{h_t} \overline u_x r \d r$.
The right-hand side accords with our intuitive understanding
of entrainment across a physically defined interface. Similarly, defining a specific momentum flux $M=2\int_0^{h_t} \overline u_x^2 r \d r$, 
the \emph{integral} or \emph{top-hat} width $h=Q/\sqrt{M}$ of an unsteady jet or plume obeys \citep{Craske2016}
\begin{equation}
  \frac{1}{\gamma}\pd{h^{2}}{t} + \pd{Q}{x}=2\alpha M^{1/2},
\label{eq:uplume_g}
\end{equation}
where $\alpha$ is an entrainment coefficient that
depends on dimensionless properties of the flow, such as the
Richardson number, dimensionless steamwise gradients and
parameters characterising the flow's radial dependence. The
dimensionless parameter $\gamma$ characterises the shape of the
mean velocity in the plume as an integral of the mean flux of
streamwise kinetic energy divided by $M^{2}/Q$. 
If one assumes self-similarity by introducing  a similarity variable $\eta=r/h$, it directly follows that 1) $h_t = \eta_t h$ where $\eta_t$ is a constant; and 2) that $\gamma$ is a constant (4/3 for a Gaussian profile). In this case, the terms in \eqref{eq:uplume_l} and \eqref{eq:uplume_g} can be matched individually with result
\begin{equation}
    \eta_t = \gamma^{-1/2}, \quad\quad 
  \alpha =-\frac{1}{2\pi M^{1/2}}\oint_{\partial \overline{\Omega}} \frac{\mybar{V}_{g}}{|\vec N_{g\perp}|} ~ \d \ell\equiv \alpha_{0}.
           \label{eq:halpha}
\end{equation}
Equation \eqref{eq:halpha} shows that for an unsteady flow that remains self-similar, the global entrainment coefficient $\alpha$ represents physical entrainment across its boundary $h_t$, entirely consistent with its classical interpretation.

However, in the vicinity of abrupt changes in the streamwise direction, unsteady jets and plumes depart significantly from self-similarity \citep{Craske2015,Craske2016}.
In this case, the equivalence between the individual terms in \eqref{eq:uplume_l} and \eqref{eq:uplume_g} is lost. 
Consequently, the strongest statement that can be made regarding the entrainment coefficient is that
\begin{equation}
\alpha=\underbrace{-\frac{1}{2\pi\,M^{1/2}}\oint_{\partial
  \overline{\Omega}} \frac{\mybar{V}_{g}}{|\vec N_{g\perp}|}
~ \d \ell}_{\alpha_{0}}+\underbrace{\frac{1}{2M^{1/2}} \left(\pd{}{t}\left(\frac{h^2}{\gamma}-h_{t}^2\right)+\frac{h^2}{ \gamma^2}\pd{\gamma}{t} \right)}_{\alpha_{1}}.
\label{eq:alpha_st}
\end{equation}
The entrainment coefficient $\alpha_{0}$ continues to account for
fluid entrained across the TNTI and therefore has a direct
physical interpretation. In contrast, the pseudo entrainment
described by $\alpha_{1}$ reconciles the definition of $\alpha$,
as stated in \eqref{eq:uplume_g} terms of $Q$ and $M$, with
entrainment across the TNTI during departures from
self-similarity. It accounts for differences between temporal
changes in the widths $h$ and $h_{t}$, in addition to temporal
changes in the parameter $\gamma$, which accounts for a change in
the shape of the mean velocity profile.

If, in view of such difficulties, one is tempted to suggest that we should abandon \eqref{eq:uplume_g} and focus on \eqref{eq:uplume_l} instead, it should be noted that \eqref{eq:uplume_g}, unlike
\eqref{eq:uplume_l}, can be readily augmented with a conservation
equation for momentum containing $\partial_{t} Q+\partial_{x}M$ to
produce a tractable model \citep{Craske2016}. Indeed, it is for this reason that establishing the connection between the local and global perspectives of entrainment is crucial.

\section{Conclusions}
\label{sec:conclusions}

Turbulent entrainment lies at the core of many important
applications in engineering and science. This paper developed an integral description of turbulent free shear flows that develop in space and/or time.  It connects local and global descriptions of turbulent entrainment, and provides a simple and clear notation to describe the intricacies of TNTI dynamics.
The description relies on the relative velocity between the fluid and the scalar interface $V_n$. By applying this description, in which the interface is defined implicitly via the isosurface $\chi=\chi_0$, in a local manner, integral equations are obtained that explicity feature the role of local entrainment.

By using an average scalar field $\overline \chi$, an equation for the global entrainment velocity $\mybar V_g$ was obtained, which resulted in equation \eqref{eq:Vnavh} formulated in terms of the interface thickness $h_t$.  For slender flows, this equation simplified to $\mybar V_g = \overline u_\zeta - \overline D h_t / \overline D t$. The associated integral equations make a statement about global entrainment.


The description can be used to provide insight into \rev{the different entrainment mechanisms} of canonical free shear flows.
One important example where it can provide \rev{insight is the parameterisation of entrainment} for plumes in crossflows. 
This flow is interesting since it will have significant contributions from both direct entrainment ($\overline u_\zeta$) and from the Leibniz terms \citep{Schatzman1978, Davidson1986}.
A detailed investigation using direct numerical simulation is currently underway that investigates both types \rev{of entrainment}.
The method will also be of interest to studying entrainment in clouds \citep{deRooy2013}.
\rev{Furthermore, the description can be applied to turbulent boundary layers \citep{Townsend1976, Chauhan2014} and their control \citep{GadElHak1991}, particularly in combination with recently developed  decomposition techniques for local \citep{Holzner2011} and global \citep{vanReeuwijk2015} descriptions of turbulent entrainment. }
The description can also be used to link entrainment to non-equilibruim turbulence
\citep{Zhou2017, Cafiero2019}.

\rev{Identifying discrete events that are responsible for entrainment has been a major focus in entrainment research since its inception \citep[e.g.][]{Townsend1976, Fernando1991}. 
The description in its current form is not directly able to link turbulent entrainment to local and discrete events, since the integrals sum over all events. 
However, it is possible to combine the entrainment descriptions to a method to identify coherent structures \citep[e.g.][]{NeamtuHalic2020}, although one should keep in mind that the choice of local or global entrainment description can isolate different aspects of entrainment (engulfing or nibbling, for example) which one can then link to each other via local-global approximate balances such as the one discussed in \S \ref{sec:localglobal}. It is hoped that the current description will be able to assist in connecting macro-scale to micro-scale entrainment events and processes.
}

\section*{Acknowledgements}

M.v.R.\ and J.C.V.\ acknowledge financial support from the H2020 Innovative Training Network COMPLETE (grant agreement no 675675). M.v.R. was additionally supported by the EPSRC project Multi-scale Dynamics at the Turbulent/Non-turbulent Interface of Jets and Plumes (grant number EP/R043175/1).
J.C.\ gratefully acknowledges an Imperial College Junior Research Fellowship award.

\section*{Declaration of interests}
The authors report no conflict of interest.
\appendix

\section{Integral identities}
\label{sec:intop}

In this appendix we derive the integral identities
\eqref{eq:gradint}-\eqref{eq:matderint} by considering volume and time
integrals over infinitesimal slices of size $\delta x\rightarrow 0$
and $\delta t\rightarrow 0$, respectively. The identity
\eqref{eq:gradint} for the integral gradient operator can be obtained
directly from \eqref{eq:divint} by substituting $\vec F = \phi \vec
e_i$ for $i \in \{x,y,z\}$; we therefore only need to prove
\eqref{eq:divint} and \eqref{eq:matderint}. We start with the proof of
\eqref{eq:divint} which is a generalisation of the method introduced
by \cite{Zhou2017} in their Appendix. The first step is to decompose
$\int_{\Omega} \div \vec F \d A$ as follows:
\begin{equation}
\int_{\Omega} \div \vec F \d A = \int_{\Omega}
\frac{\partial F_{x}}{\partial x} \d A+\int_{\Omega} \div_\perp
\vec F_\perp \d A = \int_{\Omega} \frac{\partial F_x}{\partial x}
\d A - \oint_{\partial \Omega} \vec F_\perp \cdot \vec n \d
\ell,
\label{step1}
\end{equation}
where use is made of Gauss's divergence theorem (cf. Stokes'
theorem in appendix \ref{sec:dg}) in the $y-z$ plane and
$\vec n = \vec{N}_\perp / |\vec N_\perp| = \grad_\perp \chi /
|\grad_\perp \chi|$ is the surface normal in the $y-z$ plane. 
Note that the minus sign in the last term of \eqref{step1} originates from the fact that $\vec n$ points \emph{into} $\Omega$ rather than outwards.

We now seek a formula for commuting $\int_{\Omega}$ and
$\partial/\partial x$ in \eqref{step1}. Note that
\begin{equation}
  \begin{split}
 \frac{\partial }{\partial x}\int_{\Omega} F_x \d A =
 \lim_{\delta x \to 0} \frac{1}{\delta x} \left( \int_{\Omega
   (x+\delta x)} F_x (x+\delta x) \d A - \int_{\Omega (x)} F_x
 (x) \d A \right) \\
 = \int_{\Omega (x)} \frac{\partial F_x}{\partial x} \d A +
 \lim_{\delta x \to 0} \frac{1}{\delta x} \left[ \int_{\Omega
   (x+\delta x)} F_x (x+\delta x) \d A - \int_{\Omega (x)} F_x
 (x + \delta x) \d A \right].
 \end{split}
\label{step2b}
\end{equation}
The bracketed term in \eqref{step2b} is the difference between the
surface integrals of $F_x (x+\delta x)$ over $\Omega (x+\delta
x)$ and over $\Omega (x)$ respectively. 
This integral is crucially related to the slope of the interface with the $x-$direction, $N_x / |\vec N_\perp|$.
Indeed, the amount of substance flowing into $\Omega$ at a certain location on the interface due to the slope is equal to $F_x \delta h_t \delta \ell$, where $\delta h_t$ is the change in the interface position in the $y-z$ plane (normal to $\delta \ell$) over the streamwise distance $\delta x$.
Since $\delta h_t = N_x /|\vec N_\perp| \delta x$ (Figure \ref{fig:local}b), it follows that the difference between the two surface integrals is, to leading order, equal to the curvilinear integral $\oint_{\partial \Omega} F_x (x + \delta x) N_x /|\vec N_\perp| \delta x
\d \ell$.
Hence, \eqref{step2b} becomes
\begin{equation}
\frac{\partial }{\partial x}\int_{\Omega} F_x \d A =
\int_{\Omega} \frac{\partial F_x}{\partial x} \d A +
\oint_{\partial \Omega} F_x  \frac{N_x}{|\vec N_\perp|}\d \ell.
\label{step3}
\end{equation}
Combining \eqref{step3} with \eqref{step1} leaves us with our first
main general result, identity \eqref{eq:divint}.

In the case where $\vec F = Q \vec u$ for some field $Q$, we have $F_x
(x) {N_x}/{|\vec N_\perp|} = Q(x) u_x (x) {N_x}/{|\vec
  N_\perp|}$. Defining $\d t$ to be the time required for a fluid
element to move a distance $\d x = u_x \d t$ in the streamwise
direction, \cite{Zhou2017} defined the pseudo-velocity ${\d
  h_t}/{\d t}$ which they termed $V_n$ (not to be confused with the
definition of $V_n$ in the present paper). Given that their $V_n$
equals $u_x (x) {N_x}/{|\vec N_\perp|}$, $\oint_{\partial
  \Omega} F_x (x) {N_x}/{|\vec N_\perp|}\d \ell =
\oint_{\partial \Omega} Q(x) V_n {N_x}/{|\vec N_\perp|}\d
\ell$ in terms of their pseudo-velocity $V_n$ (see also section B.2 of
Appendix B). This establishes the correspondence between the results
in their Appendix (which they gave for $Q=1$) and \eqref{step3},
\eqref{eq:divint} here.

We now proceed with the proof of identity
\eqref{eq:matderint}. Integrating the material derivative over $\Omega$ yields
\begin{equation}
\int_{\Omega} \frac{D \phi}{D t} \d A = \int_{\Omega}
\frac{\partial }{\partial t} \phi ~ \d A + \int_{\Omega} \div
(\phi \vec u ) ~ \d A  - \int_{\Omega} \phi \div \vec u  \d A
\end{equation}
and then making use of \eqref{eq:divint} for $\vec F = \phi \vec u$,
\begin{equation}
\int_{\Omega} \frac{D \phi}{D t} \d A = \int_{\Omega}
\frac{\partial }{\partial t} \phi ~ \d A + \frac{\partial }{\partial
  x}\int_{\Omega} \phi u_x \d A - \oint_{\partial \Omega}
\phi \vec u \cdot \frac{\vec N}{| \vec N_\perp|} \d \ell   - \int_{\Omega} \phi \div \vec u  \d A.
\label{step4}
\end{equation}
Following \eqref{step2b}, we now wish to commute
$\int_{\Omega}$ and $\partial/\partial t$:
\begin{equation}
\begin{split}
 \frac{\partial }{\partial t}\int_{\Omega} \phi \d A =
 \lim_{\delta t \to 0} \frac{1}{\delta t} \left( \int_{\Omega
   (t+\delta t)} \phi (t+\delta t) \d A - \int_{\Omega (t)} \phi
 (t) \d A \right) \\
 = \int_{\Omega (t)} \frac{\partial \phi}{\partial t} \d A +
 \lim_{\delta t \to 0} \frac{1}{\delta t} \left[ \int_{\Omega
   (t+\delta t)} \phi (t+\delta t) \d A - \int_{\Omega (t)} \phi
 (t + \delta t) \d A \right].
 \end{split}
\label{step5b}
\end{equation}
The bracketed term in \eqref{step5b} is the difference between the
surface integrals of $\phi (t+\delta t)$ over $\Omega
(t+\delta t)$ and over $\Omega (t)$ respectively. 
The interface, which moves at normal velocity $v_n = \vec v \cdot \vec N$, will move at velocity $v_n / | \vec N_\perp|$ when projected onto the $y-z$ plane.
Thus, over a time increment $\delta t$, the interface element of length $\delta \ell$ will sweep an area in the $y-z$ plane equal to $-v_n/ | \vec N_\perp| \delta t \delta \ell$ (the minus sign once more originates from the inward pointing normal $\vec N$).
This implies that difference between the two surface integrals in \eqref{step5b} is, to leading order, equal to the curvilinear integral $-\oint_{\partial \Omega} \phi
(t+\delta t) v_n/|\vec N_\perp|\delta t \d \ell$.
Hence, \eqref{step5b} becomes
\begin{equation}
\frac{\partial }{\partial t}\int_{\Omega} \phi \d A =
\int_{\Omega} \frac{\partial \phi}{\partial t} \d A -
\oint_{\partial \Omega} \phi \frac{v_{n}}{|\vec N_\perp|}\d {\ell}.
\label{step6}
\end{equation}
Combining \eqref{step6} with \eqref{step4} and invoking the relative
iso-surface velocity $\vec V = \vec v - \vec u$, leaves us with our
second main general result, identity \eqref{eq:matderint}, i.e.
\begin{equation*}
\int_{\Omega} \frac{D \phi}{D t} \d A = 
\frac{\partial }{\partial t}\int_{\Omega} \phi ~ \d A  +    \frac{\partial}{\partial x} \int_{\Omega} u_x \phi ~ \d A +
   \oint_{\partial \Omega}  \frac{V_{n}}{| \vec N_\perp|} \phi ~ \d \ell - \int_{\Omega} \phi \div \vec u  \d A.
 \end{equation*}
%
Noting that the approaches used in \eqref{step2b} and \eqref{step5b}
are equivalent and account for the commutation of integration with
differentiation with respect to either time or space, we abstract
and generalise our results using differential geometry in the
following section.

\section{Differential geometry}
\label{sec:dg}


In this appendix we regard the region $\Omega$, defined by an
isosurface of $\chi$, as a submanifold whose shape changes as a
function of \emph{codimensions} $x$ and $t$, for example. In
manipulating integrals of partial derivatives over $\Omega$, one is
faced with two distinct types of expression. The first involves
derivatives in directions that lie within the dimensions of
$\Omega$ and the second involves derivatives in directions that
lie in the codimension of $\Omega$. The first can be manipulated
using a generalised version of Stokes' theorem, whilst the second
require a generalised form of Leibniz's rule for commuting
integration and partial differentiation. The first involve
physical fluxes and velocities at the boundary
$\partial\Omega$. The second, in contrast, involve pseudo fluxes
and velocities that account for deformations of the integration
domain with respect to changes in a given codimension. Since the
integration domain is specified by $\chi$ independently of
physical boundary fluxes, the two are independent.

\begin{figure}
  \begin{center}
    \includegraphics{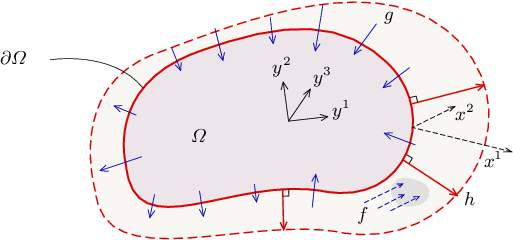}
  \end{center}
  \caption{A submanifold $\Omega$ described by local coordinates
    $y^{1},\ldots,y^{n}$ and parameterised over codimensions
    $x^{1},\ldots,x^{m}$. The physical flux in the plane of
    $\Omega$ corresponds to the $(m+n-1)$-form $g$ and the pseudo
    flux, arising from the normal flux $f$ through deformations in
    $\partial\Omega$ as one moves with unit `velocity' in the
    direction $x^{i}$ corresponds to the $(m+n-1)$-form $h$.}
  \label{diag:submanifold}
\end{figure}

To crystallise these ideas, consider an $n-$dimensional slice
$\Omega$ through an $N-$dimensional manifold defined by
$\chi(\vec{x},\vec{y})\geq\chi_{0}$ by fixing $m=N-n$ codimensions
$\vec{x}$, as depicted in figure \ref{diag:submanifold}. The slice
itself can be traversed locally by coordinates
$y^{1},\ldots,y^{n}$. Components of a differential $(N-1)$-form
$\omega$ can be partitioned into fluxes $f$ and $g$ that are
either normal or tangential to the area form
$\d A = \d y^{1}\wedge\d y^{2}\wedge\ldots\wedge\d y^{n}$,
respectively. For example, if $x^{1}=x$, $y^{1}=y$ and $y^{2}=z$,
a flux $\omega$ can be partitioned, such that:

\begin{equation}
  \omega = \underbrace{\omega_{x}\d y\wedge \d z}_{f}+\underbrace{\omega_{y}\d z\wedge \d x+\omega_{z}\d x\wedge \d y}_{g}.
  \label{eq:dgflux}
\end{equation}
Integrals of the $N$-form $\d g$ at a given point in the
codomain can be evaluated using a generalised version of the
fundamental theorem of calculus in the form of Stokes'
theorem. With slight abuse of notation, because $\d g$ is an
$N$-form rather than an $n$-form, we express the application of
Stokes' theorem over the slice $\Omega$ as

\begin{equation}
  \int_{\Omega}\mathrm{d}g=\int_{\partial\Omega}g,
  \label{eq:stokes}
\end{equation}
which we will refer to as a \emph{partial integral}
\citep{WhiHboo1957a} that results in a differential form
containing, in the case of \eqref{eq:stokes},
$\d x^{1}\wedge\ldots\wedge\d x^{m}$. A more rigorous treatment
of the operation would consider integrals along a \emph{fibre}
($\Omega$) of a \emph{fibre bundle} (the entire space). In either
case, equation \eqref{eq:stokes} states that integrals of
derivatives with respect to $y^{1}\,\ldots,y^{n}$ can be evaluated
as surface integrals that account for boundary transport.

The manipulation $\d f$ is fundamentally different from $\d g$
because $\d f$, unlike $\d g$, involves partial derivatives with
respect to the codimensions $x^{1},\ldots,x^{m}$. Consequently,
partial integrals of $\d f$ over $\Omega$ satisfy a generalised
version of Leibniz's rule that specifies how to commute
integration with exterior differentiation:

\begin{equation}
\int_{\Omega}\d f= \d\int_{\Omega}f -\int_{\partial\Omega}h,
  \label{eq:commute}
\end{equation}
in which the partial integral of $f$ over $\Omega$ produces an
$(m-1)$-form to which the exterior derivative $\mathrm{d}$ can be
applied. The final term in \eqref{eq:commute} contains $h$, which
is a differential $(N-1)$-form corresponding to a \emph{pseudo}
flux, in contrast with the physical flux $g$. The \emph{pseudo}
flux $h$ results from $f$ and the dependence of $\Omega$ on the
codimensions, and therefore depends on the geometry of the surface
$\chi-\chi_{0}=0$.

To determine $h$, we define a vector $Y$ that is tangent to
$\Omega$, is perpendicular to an isosurface of $\chi$:

\begin{equation}
  (\grad\chi\intprod\d A)(Y)=0,
  \label{eq:dgperp}
\end{equation}
and corresponds to a unit rate of change down $\chi$, such that
$\d\chi(Y)=-1$. Consequently, the vector $(\partial_{x^{i}}\chi)Y$
acts in a direction that is perpendicular to $\partial\Omega$ and
describes the change in $\partial\Omega$ that occurs as one moves
with unit `velocity' along the $i^{\mathrm{th}}$ codimension. The
effect of changes in $\Omega$ on integrals is captured by Cartan's
magic formula for the Lie derivative
$X\intprod\d\omega=\mathcal{L}_{X}\omega-\d(X\intprod\omega)$,
where $\mathcal{L}_{X}\omega$ describes the change in $\omega$
along the flow defined by $X$. Setting
$X=\partial_{x^{i}}+(\partial_{x^{i}}\chi)Y$ and focusing on each
component $f_{i}$ of $f$ in \eqref{eq:dgflux}, leads to a
generalised version of Leibniz's rule \citep[see, for
example,][]{FlaHamm1973a} for partial integration over $\Omega$:

\begin{equation}
  \int_{\Omega}\pd{}{x^{i}}\intprod\d f_{i}
  = \pd{}{x^{i}}\int_{\Omega}f_{i}-\int_{\partial\Omega}\pd{\chi}{x^{i}}Y\intprod f_{i},
  \label{eq:leibniz}
\end{equation}
which is an $(m-1)$-form. Applying $\d x^{i}\wedge$ to
\eqref{eq:leibniz}, and summing over each codimension $i$, shows
that each term in \eqref{eq:leibniz} corresponds to the respective
terms in \eqref{eq:commute} and therefore reveals that
\begin{equation}
  h = Y\intprod (\d\chi\wedge f),
  \label{eq:dgh}
\end{equation}
which is fundamental in determining the boundary contribution of
fluxes that are perpendicular to $\d A$. Combining \eqref{eq:dgh}
with \eqref{eq:stokes} and \eqref{eq:commute} leads to a general
formula for commuting exterior differentiation and partial
integration over a submanifold:

\begin{equation}
  \int_{\Omega}\d\omega-\d\int_{\Omega}\omega=\underbrace{\int_{\partial{\Omega}}g}_{\text{Stokes}}
  -\underbrace{\int_{\partial\Omega}h}_{\text{Leibniz}}.
  \label{eq:dgcomb}
\end{equation}
As illustrated in figure \ref{diag:submanifold}, the commutation
of integration and exterior differentiation leads to physical
fluxes $g$ in the plane of $\d A$ in addition to \emph{pseudo}
fluxes $h$, which account for the flow $f$ through the contraction
and expansion of $\Omega$ as the codimensions $t$ and $x$ change.

\subsection{Example: integration of $D\phi/D t$ over $\Omega(x,t)\ni (y,z)$}
\label{sec:dgexample}

To integrate $D\phi/Dt$ over an area $\Omega(x,t)$, we identify
$x^{1}=t$, $x^{2}=x$ as the submanifold's codimensions and
$y^{1}=y$, $y^{2}=z$ as the submanifold's coordinates. We
decompose $D\phi/Dt$ to incorporate the divergence of a flux
$\omega$ and indicate the correspondance with
$\d\omega=\d f+\d g$:

\begin{equation}
  \frac{D\phi}{D t}=\underbrace{\pd{\phi}{t} + \pd{u_{x}\phi}{x}}_{\d f}+\underbrace{\pd{u_{y}\phi}{y}+\pd{u_{z}\phi}{z}}_{\d g}-\phi\grad\cdot\vec{u},
\end{equation}
which, regarding $D\phi/Dt$ as a differential 4-form, implies that 
\begin{equation}
  \omega = \underbrace{\phi\,\d x\wedge\d y\wedge\d z
           -u_{x}\phi\,\d t\wedge\d y\wedge\d z}_{f}
         +\underbrace{u_{y}\phi\,\d t\wedge\d x\wedge\d z
           -u_{z}\phi\,\d t\wedge\d x\wedge\d y}_{g}.
\end{equation}
Here, the tangent vector $Y$ is:
\begin{equation}
  Y=Y_{y}\pd{}{y}+Y_{z}\pd{}{z}=-\left(\left(\pd{\chi}{y}\right)^{2}+\left(\pd{\chi}{z}\right)^{2}\right)^{-1}
\left(\pd{\chi}{y}\pd{}{y} + \pd{\chi}{z}\pd{}{z}\right),
\label{eq:Yexample}
\end{equation}
which, using \eqref{eq:dgcomb} and omitting the $\d t\wedge\d x$
that remains after partial integration, results in
\begin{equation}
  \begin{split}
  \int_{\Omega}\frac{D\phi}{Dt}\d A= \pd{}{t}&\int_{\Omega}\phi\d A+\pd{}{x}\int_{\Omega} u_{x}\phi\d A+\int_{\partial\Omega}\left(Y_{z}\pd{\chi}{t}+u_{x}Y_{z}\pd{\chi}{x}-u_{z}\right)\phi\d y\\
  &  -\int_{\partial\Omega}\left(Y_{y}\pd{\chi}{t}+u_{x}Y_{y}\pd{\chi}{x}-u_{y}\right)\phi\d z -\int_{\Omega}\phi\grad\cdot\vec{u}\d A.
\end{split}
\label{eq:dphidt}
\end{equation}
Identities for the integral of $\grad\phi$ and the area of
$\Omega$ can be obtained as corollaries of \eqref{eq:dphidt} by
substitution of $\vec{u}=(1,1,1)^{T}$ and $\phi=1$, respectively.

\subsection{Summary and connection with appendix \ref{sec:intop}}

Our results indicate that the commutation of integration and
differentiation leads to storage terms and \emph{pseudo} boundary
fluxes that depend on the $t$ and $x$ dependence of the domain of
integration. Such fluxes are distinct from the physical boundary
fluxes that are obtained by applying Stokes' theorem to the
divergence of fluxes that are in the same plane as the domain of
integration. 

To link \eqref{eq:dgcomb} with appendix \ref{sec:intop} it is
necessary to note that the vector $Y$ introduced at
\eqref{eq:dgperp}, and written explicitly in \eqref{eq:Yexample},
corresponds to $-\vec{n}/|\grad_{\perp}\chi|$ and
that

\begin{equation}
  \pd{\chi}{t}=-v_{n}|\grad\chi|,\quad\quad \pd{\chi}{x}=u_{x}N_{x}|\grad\chi|.
  \label{eq:dchidt}
\end{equation}

\noindent As made explicit in the fourth term of
\eqref{eq:dphidt}, each term in the integral of $h$ accounts for
$Y$, which is normal to $\partial\Omega$ according to \eqref{eq:dgperp}, scaled by
either $\partial_{t}\chi$ or $\partial_{x}\chi$. Consequently, by
expanding \eqref{eq:matderint} for $\phi=1$, using
\eqref{eq:dchidt} and recalling that
$|\vec N_{\perp}|=|\grad_{\perp}\chi|/|\grad\chi|$,
\begin{equation}
      \frac{\partial}{\partial t} \int_{\Omega} \d A + \frac{\partial}{\partial x} \int_{\Omega} u_x  \d A 
      = \oint_{\partial \Omega}\underbrace{\vec{u}\cdot\vec{n}\d \ell}_{-g}
       - \oint_{\partial \Omega}\underbrace{\frac{v_n}{|\vec N_\perp |}
       - u_x \frac{ N_x}{|\vec N_\perp |}\d \ell}_{-h},
\end{equation}
which is a special case of the relation \eqref{eq:dgcomb}. The inward
boundary flux $g=\vec{u}\cdot \vec{n}\d\ell$ results from Stokes' theorem and
is identical to $V_{f}\d\ell$ in \cite{Zhou2017}. In contrast, the
pseudo fluxes $h$ result from Leibniz's rule for commuting integration
and differentiation. The terms in $h$ correspond to perpendicular
fluxes through contractions and expansions of the boundary
$\partial\Omega$ as one traverses the \emph{codimensions} $t$
and $x$ with unit velocity (see figure \ref{diag:submanifold}). Specifically, the term $u_{x}N_{x}/|\vec{N}_{\perp}|\d\ell$ corresponds to the increase in
volume flux due to changes in the location of $\partial\Omega$
as one moves, one unit in the $x$ direction and is identical to
$V_{n}\d\ell$ in \cite{Zhou2017} (stressing, again, that this $V_n$ is
not the same as the $V_n$ defined in the present paper).

A practical advantage of the abstract approach adopted in appendix
\ref{sec:dg} is that the resulting surface integrals (see
\eqref{eq:dphidt}, for example) are given explicitly in terms of
coordinate differentials $\d y$, $\d z$. The integral expressions
can therefore be readily evaluated on a numerical grid and, since
differential forms $\d y$, $\d z$, $\d y\wedge\d z$ have an
orientation, automatically account for the orientation of a
surface and its normal.

\bibliographystyle{jfm}
\bibliography{entrainment.bib} 

\begin{thebibliography}{72}
\expandafter\ifx\csname natexlab\endcsname\relax\def\natexlab#1{#1}\fi
\def\au#1{#1} \def\ed#1{#1} \def\yr#1{#1}\def\at#1{#1}\def\jt#1{\textit{#1}}
  \def\bt#1{#1}\def\bvol#1{\textbf{#1}} \def\vol#1{#1} \def\pg#1{#1}
  \def\publ#1{#1}\def\arxiv#1{#1}\def\org#1{#1}\def\st#1{\textit{#1}}

\bibitem[Bisset {\em et~al.\/}(2002)Bisset, Hunt \& Rogers]{Bisset2002}
{\sc \au{Bisset, D.~K.}, \au{Hunt, J. C.~R.} \& \au{Rogers, M.~M.}} \yr{2002}
  \at{The turbulent/non-turbulent interface bounding a far wake}.  \jt{J. Fluid
  Mech.}  \bvol{451},  \pg{383--410}.

\bibitem[Burridge {\em et~al.\/}(2017)Burridge, Parker, Kruger, Partridge \&
  Linden]{Burridge2017}
{\sc \au{Burridge, H.~C.}, \au{Parker, D.~A.}, \au{Kruger, E.~S.},
  \au{Partridge, J.~L.} \& \au{Linden, P.~F.}} \yr{2017}  \at{Conditional
  sampling of a high {P}\'eclet number turbulent plume and the implications for
  entrainment}.  \jt{J. Fluid Mech.}  \bvol{823},  \pg{26--56}.

\bibitem[Cafiero \& Vassilicos(2019)]{Cafiero2019}
{\sc \au{Cafiero, G.} \& \au{Vassilicos, J.C.}} \yr{2019}  \at{Non-equilibrium
  turbulence scalings and self-similarity in turbulent planar jets}.  \jt{Proc.
  R. Soc. Lond. A}  \bvol{475},  \pg{20190038}.

\bibitem[Cantwell \& Coles(1983)]{Cantwell1983}
{\sc \au{Cantwell, B.} \& \au{Coles, D.}} \yr{1983}  \at{An experimental study
  of entrainment and transport in the turbulent near wake of a circular
  cylinder}.  \jt{J. Fluid Mech.}  \bvol{136},  \pg{321--374}.

\bibitem[Chauhan {\em et~al.\/}(2014)Chauhan, Philip, de~Silva, Hutchins \&
  Marusic]{Chauhan2014}
{\sc \au{Chauhan, K.}, \au{Philip, J.}, \au{de~Silva, C.~M.}, \au{Hutchins, N.}
  \& \au{Marusic, I.}} \yr{2014}  \at{The turbulent/non-turbulent interface and
  entrainment in a boundary layer}.  \jt{J. Fluid Mech.}  \bvol{742},
  \pg{119--151}.

\bibitem[Ching {\em et~al.\/}(1995)Ching, Fernando \& Robles]{Ching1995}
{\sc \au{Ching, C.Y.}, \au{Fernando, H.J.S.} \& \au{Robles, A.}} \yr{1995}
  \at{Break-down of line plumes in turbulent environments}.  \jt{Journal of
  Geophysical Research: Oceans}  \bvol{100},  \pg{4707--4713}.

\bibitem[Corrsin \& Kistler(1955)]{Corrsin1955}
{\sc \au{Corrsin, S.} \& \au{Kistler, A.L}} \yr{1955}  \bt{Free stream
  boundaries of turbulent flows}. {\em Tech. Rep.\/} 1244.  \org{NACA}.

\bibitem[Craske \& van Reeuwijk(2015)]{Craske2015}
{\sc \au{Craske, J.} \& \au{van Reeuwijk, M.}} \yr{2015}  \at{Energy dispersion
  in turbulent jets. {P}art 1. {D}irect simulation of steady and unsteady
  jets}.  \jt{J. Fluid Mech.}  \bvol{763},  \pg{500--537.}

\bibitem[Craske \& van Reeuwijk(2016)]{Craske2016}
{\sc \au{Craske, J.} \& \au{van Reeuwijk, M.}} \yr{2016}  \at{Generalised
  unsteady plume theory}.  \jt{J. Fluid Mech.}  \bvol{792},  \pg{1013--1052.}

\bibitem[Da~Silva {\em et~al.\/}(2014)Da~Silva, Hunt, Eames \&
  Westerweel]{DaSilva2014}
{\sc \au{Da~Silva, C.~B.}, \au{Hunt, J. C.~R.}, \au{Eames, I.} \&
  \au{Westerweel, J.}} \yr{2014}  \at{Interfacial layers between regions of
  different turbulence intensity}.  \jt{Annu. Rev. Fluid Mech.}  \bvol{46},
  \pg{567--590}.

\bibitem[Da~Silva \& M\'{e}tais(2002)]{DaSilva2002}
{\sc \au{Da~Silva, C.~B.} \& \au{M\'{e}tais, O.}} \yr{2002}  \at{On the
  influence of coherent structures upon interscale interactions in turbulent
  plane jets}.  \jt{J. Fluid Mech.}  \bvol{473},  \pg{103--145}.

\bibitem[Da~Silva \& Pereira(2008)]{DaSilva2008}
{\sc \au{Da~Silva, C.~B.} \& \au{Pereira, J. C.~F.}} \yr{2008}  \at{Invariants
  of the velocity-gradient, rate-of-strain, and rate-of-rotation tensors across
  the turbulent/nonturbulent interface in jets}.  \jt{Phys. Fluids}
  \bvol{20}~(5),  \pg{055101}.

\bibitem[Davidson(1986)]{Davidson1986}
{\sc \au{Davidson, G.A.}} \yr{1986}  \at{A discussion of {S}chatzmann's
  integral plume nodel from a control volume viewpoint}.  \jt{J. Clim. Appl.
  Met.}  \bvol{25},  \pg{858--866}.

\bibitem[De~Wit {\em et~al.\/}(2014)De~Wit, Van~Rhee \& Keetels]{DeWit2014}
{\sc \au{De~Wit, L.}, \au{Van~Rhee, C.} \& \au{Keetels, G.}} \yr{2014}
  \at{Turbulent interaction of a buoyant jet with cross-flow}.  \jt{J. Hydr.
  Eng.}  \bvol{140}~(12),  \pg{04014060}.

\bibitem[Deardorff {\em et~al.\/}(1980)Deardorff, Willis \&
  Stockton]{Deardorff1980}
{\sc \au{Deardorff, J.~W.}, \au{Willis, G.~E.} \& \au{Stockton, B.~H.}}
  \yr{1980}  \at{Laboratory studies of the entrainment zone of a convectively
  mixed layer.}  \jt{J. Fluid Mech.}  \bvol{100},  \pg{41--64.}

\bibitem[Devenish {\em et~al.\/}(2010)Devenish, Rooney \&
  Thomson]{Devenish2010}
{\sc \au{Devenish, B.J.}, \au{Rooney, G.G.} \& \au{Thomson, D.J.}} \yr{2010}
  \at{Large-eddy simulation of a buoyant plume in uniform and stably stratified
  environments}.  \jt{J. Fluid Mech.}  \bvol{652},  \pg{75 -- 103}.

\bibitem[Dopazo {\em et~al.\/}(2007)Dopazo, Mart\'in \& Hierro]{Dopazo2007}
{\sc \au{Dopazo, C.}, \au{Mart\'in, J.} \& \au{Hierro, J.}} \yr{2007}
  \at{Local geometry of isoscalar surfaces}.  \jt{Physical Review E}
  \bvol{76}~(5),  \pg{056316}.

\bibitem[Fernando(1991)]{Fernando1991}
{\sc \au{Fernando, H. J.~S.}} \yr{1991}  \at{Turbulent mixing in stratified
  fluids}.  \jt{Annu. Rev. Fluid Mech.}  \bvol{23},  \pg{455--493}.

\bibitem[Flanders(1973)]{FlaHamm1973a}
{\sc \au{Flanders, Harley}} \yr{1973}  \at{Differentiation under the integral
  sign}.  \jt{American Math. Monthly}  \bvol{80}~(6),  \pg{615--627}.

\bibitem[Gad-El-Hak \& Bushnell(1991)]{GadElHak1991}
{\sc \au{Gad-El-Hak, M.} \& \au{Bushnell, D.~M.}} \yr{1991}  \at{Separation
  control: review}.  \jt{Journal of Fluids Engineering}  \bvol{115},
  \pg{5--30}.

\bibitem[Garcia \& Mellado(2014)]{Garcia2014}
{\sc \au{Garcia, J.R.} \& \au{Mellado, J.P.}} \yr{2014}  \at{The two-layer
  structure of the entrainment zone in the convective boundary layer}.  \jt{J.
  Atmos. Sci.}  \bvol{71},  \pg{1935--1955}.

\bibitem[Gaskin {\em et~al.\/}(2004)Gaskin, McKernan \& Xue]{Gaskin2004}
{\sc \au{Gaskin, S.J.}, \au{McKernan, M.} \& \au{Xue, F.}} \yr{2004}  \at{The
  effect of background turbulence on jet entrainment: an experimental study of
  a plane jet in a shallow coflow}.  \jt{Journalof Hydraulic Research}
  \bvol{42}~(5),  \pg{533--542.}

\bibitem[Head(1958)]{Head1958}
{\sc \au{Head, M.R.}} \yr{1958}  \bt{Entrainment in the turbulent boundary
  layer}. Reports \& Memoranda 3152.  \org{Ministry of Aviation}.

\bibitem[Holzner \& Luethi(2011)]{Holzner2011}
{\sc \au{Holzner, M.} \& \au{Luethi, B.}} \yr{2011}  \at{Laminar superlayer at
  the turbulence boundary}.  \jt{Phys. Rev. Lett.}  \bvol{106}~(13),
  \pg{134503}.

\bibitem[Holzner \& van Reeuwijk(2017)]{Holzner2017}
{\sc \au{Holzner, M.} \& \au{van Reeuwijk, M.}} \yr{2017}  \at{The
  turbulent/nonturbulent interface in penetrative convection}.  \jt{J. Turbul.}
   \bvol{18},  \pg{260--270}.

\bibitem[Hunt \& Burridge(2015)]{Hunt2015}
{\sc \au{Hunt, G.R.} \& \au{Burridge, H.C.}} \yr{2015}  \at{Fountains in
  industry and nature}.  \jt{Annu. Rev. Fluid Mech.}  \bvol{47},
  \pg{195--220}.

\bibitem[Hunt {\em et~al.\/}(1983)Hunt, Rottman \& Britter]{Hunt1983}
{\sc \au{Hunt, J. C.~R.}, \au{Rottman, J.W.} \& \au{Britter, R.E.}} \yr{1983}
  Some physical processes involved in the dispersion of dense gases.  \bt{In
  {\em Proc. {UITAM} Symp. on Atmospheric dispersion of heavy gases and small
  particles\/} (ed. \ed{G.~Ooms \& H.~Tennekes})},  \pg{pp. 361--395}.
  \publ{Springer}.

\bibitem[Hussein {\em et~al.\/}(1994)Hussein, Capp \& George]{Hussein1994}
{\sc \au{Hussein, H.~J.}, \au{Capp, S.~P.} \& \au{George, W.~K.}} \yr{1994}
  \at{Velocity measurements in a high-{R}eynolds number, momentum-conserving,
  axisymmetric, turbulent jet.}  \jt{J. Fluid Mech.}  \bvol{258},  \pg{31--75.}

\bibitem[Jahanbakhshi \& Madnia(2018)]{Jahanbakhshi2018}
{\sc \au{Jahanbakhshi, R.} \& \au{Madnia, C.}} \yr{2018}  \at{The effect of
  heat release on the entrainment in a turbulent mixing layer.}  \jt{J. Fluid
  Mech.}  \bvol{844}.

\bibitem[Jonker {\em et~al.\/}(2013)Jonker, Van~Reeuwijk, Sullivan \&
  Patton]{Jonker2013}
{\sc \au{Jonker, H.J.J.}, \au{Van~Reeuwijk, M.}, \au{Sullivan, P.} \&
  \au{Patton, E.}} \yr{2013}  \at{On the scaling of shear-driven entrainment: A
  {DNS} study.}  \jt{J Fluid Mech.}  \bvol{732},  \pg{150--165}.

\bibitem[Kankanwadi \& Buxton(2019)]{Kankanwadi2019}
{\sc \au{Kankanwadi, K.} \& \au{Buxton, O.}} \yr{2019} Turbulent entrainment
  from a turbulent background.  \bt{In {\em Eleventh International Symposium on
  Turbulence and Shear Flow Phenomena\/}}.

\bibitem[Kato \& Phillips(1969)]{Kato1969}
{\sc \au{Kato, H.} \& \au{Phillips, O.~M.}} \yr{1969}  \at{On the penetration
  of a turbulent layer into stratified fluid.}  \jt{J. Fluid Mech.}  \bvol{37},
   \pg{643--655.}

\bibitem[Kotsovinos(1978)]{KotNjfm1978a}
{\sc \au{Kotsovinos, N.~E.}} \yr{1978}  \at{A note on the conservation of the
  volume flux in free turbulence}.  \jt{J. Fluid Mech.}  \bvol{86},
  \pg{201--203}.

\bibitem[Krug {\em et~al.\/}(2017)Krug, Chung, Philip \& Marusic]{Krug2017}
{\sc \au{Krug, D.}, \au{Chung, D.}, \au{Philip, J.} \& \au{Marusic, I.}}
  \yr{2017}  \at{Global and local aspects of entrainment in temporal plumes}.
  \jt{J. Fluid Mech.}  \bvol{812},  \pg{222--250}.

\bibitem[Krug {\em et~al.\/}(2015)Krug, Holzner, Luethi, Wolf, Kinzelbach \&
  Tsinober]{Krug2015}
{\sc \au{Krug, D.}, \au{Holzner, M.}, \au{Luethi, B.}, \au{Wolf, M.},
  \au{Kinzelbach, W.} \& \au{Tsinober, A.}} \yr{2015}  \at{The
  turbulent/non-turbulent interface in an inclined dense gravity current}.
  \jt{J. Fluid Mech.}  \bvol{765},  \pg{303--324}.

\bibitem[List(1982)]{LisEafm1982a}
{\sc \au{List, E.~J.}} \yr{1982}  \at{Turbulent jets and plumes}.  \jt{Annu.
  Rev. Fluid Mech.}  \bvol{14}~(1),  \pg{189--212}.

\bibitem[Mahesh(2013)]{Mahesh2013}
{\sc \au{Mahesh, K.}} \yr{2013}  \at{The interaction of jets with cross-flow}.
  \jt{Annu. Rev. Fluid Mech.}  \bvol{45},  \pg{379--407}.

\bibitem[Mellado(2012)]{Mellado2012}
{\sc \au{Mellado, JP.}} \yr{2012}  \at{Direct numerical simulation of free
  convection over a heated plate}.  \jt{J Fluid Mech.}  \bvol{712},
  \pg{418--450}.

\bibitem[Mellado(2017)]{Mellado2017}
{\sc \au{Mellado, J.~P.}} \yr{2017}  \at{Cloud-top entrainment in stratocumulus
  clouds}.  \jt{Annu. Rev. Fluid Mech.}  \bvol{49}~(1),  \pg{145--169}.

\bibitem[Neamtu-Halic {\em et~al.\/}(2020)Neamtu-Halic, Krug, Mollicone, van
  Reeuwijk, Haller \& Holzner]{NeamtuHalic2020}
{\sc \au{Neamtu-Halic, M.~M.}, \au{Krug, D.}, \au{Mollicone, J.P.}, \au{van
  Reeuwijk, M.}, \au{Haller, G.} \& \au{Holzner, M.}} \yr{2020}  \at{Connecting
  the time evolution of the turbulence interface to coherent structures}.
  \jt{Journal of Fluid Mechanics}  \bvol{898},  \pg{A3}.

\bibitem[Obligado {\em et~al.\/}(2016)Obligado, Dairay \&
  Vassilicos]{Obligado2016}
{\sc \au{Obligado, M.}, \au{Dairay, T.} \& \au{Vassilicos, J.~C.}} \yr{2016}
  \at{Nonequilibrium scalings of turbulent wakes.}  \jt{Phys. Rev. Fluids}
  \bvol{1}~(4),  \pg{044409}.

\bibitem[Odier {\em et~al.\/}(2014)Odier, Chen \& Ecke]{Odier2014}
{\sc \au{Odier, P.}, \au{Chen, J.} \& \au{Ecke, R.~E.}} \yr{2014}
  \at{Entrainment and mixing in a laboratory model of oceanic overflow}.
  \jt{J. Fluid Mech.}  \bvol{746},  \pg{498--535}.

\bibitem[Philip {\em et~al.\/}(2014)Philip, Meneveau, de~Silva \&
  Marusic]{Philip2014}
{\sc \au{Philip, J.}, \au{Meneveau, C.}, \au{de~Silva, C.~M.} \& \au{Marusic,
  I.}} \yr{2014}  \at{Multiscale analysis of fluxes at the
  turbulent/non-turbulent interface in high {R}eynolds number boundary layers}.
   \jt{Phys. Fluids}  \bvol{26},  \pg{015105}.

\bibitem[Pope(2000)]{Pope2000}
{\sc \au{Pope, S.~B.}} \yr{2000} {\em Turbulent flows\/}.  \publ{Cambridge
  University Press}.

\bibitem[Rajaratnam(1976)]{Rajaratnam1976}
{\sc \au{Rajaratnam, N.}} \yr{1976} {\em Turbulent Jets\/}. {\em Developments
  in Water Science\/} 5.  \publ{Elsevier}.

\bibitem[Redford {\em et~al.\/}(2012)Redford, Castro \& Coleman]{Redford2012}
{\sc \au{Redford, J.~A.}, \au{Castro, I.~P.} \& \au{Coleman, G.~N.}} \yr{2012}
  \at{On the universality of turbulent axisymmetric wakes}.  \jt{J. Fluid
  Mech.}  \bvol{710},  \pg{419--452}.

\bibitem[van Reeuwijk \& Craske(2015)]{vanReeuwijk2015}
{\sc \au{van Reeuwijk, M.} \& \au{Craske, J.}} \yr{2015}  \at{Energy-consistent
  entrainment relations for jets and plumes}.  \jt{J. Fluid Mech.}  \bvol{782},
   \pg{333 -- 355}.

\bibitem[van Reeuwijk \& Holzner(2014)]{vanReeuwijk2014}
{\sc \au{van Reeuwijk, M.} \& \au{Holzner, M.}} \yr{2014}  \at{The turbulence
  boundary of a temporal jet}.  \jt{J. Fluid Mech.}  \bvol{739},
  \pg{254--275}.

\bibitem[van Reeuwijk {\em et~al.\/}(2019)van Reeuwijk, Holzner \&
  Caulfield]{vanReeuwijk2019}
{\sc \au{van Reeuwijk, M.}, \au{Holzner, M.} \& \au{Caulfield, C.~P.}}
  \yr{2019}  \at{Mixing and entrainment are suppressed in inclined gravity
  currents}.  \jt{J. Fluid Mech.}  \bvol{873},  \pg{786--815.}

\bibitem[van Reeuwijk {\em et~al.\/}(2018)van Reeuwijk, Krug \&
  Holzner]{vanReeuwijk2018}
{\sc \au{van Reeuwijk, M.}, \au{Krug, D.} \& \au{Holzner, M.}} \yr{2018}
  \at{Small-scale entrainment in inclined gravity currents}.  \jt{Environ.
  Fluid Mech.}  \bvol{18}~(1),  \pg{225--239}.

\bibitem[van Reeuwijk {\em et~al.\/}(2016)van Reeuwijk, Salizzoni, Hunt \&
  Craske]{vanReeuwijk2016}
{\sc \au{van Reeuwijk, Maarten}, \au{Salizzoni, Pietro}, \au{Hunt, Gary~R.} \&
  \au{Craske, John}} \yr{2016}  \at{Turbulent transport and entrainment in jets
  and plumes: A dns study}.  \jt{Phys. Rev. Fluids}  \bvol{1},  \pg{074301}.

\bibitem[de~Rooy(2013)]{deRooy2013}
{\sc \au{de~Rooy, W.C.~\emph{et al.}}} \yr{2013}  \at{Entrainment and
  detrainment in cumulus convection: an overview}.  \jt{Quart. J. Roy. Meteor.
  Soc.}  \bvol{139}~(670),  \pg{1--19}.

\bibitem[Scase {\em et~al.\/}(2006)Scase, Caulfield, Dalziel \&
  Hunt]{ScaMjfm2006c}
{\sc \au{Scase, M.~M.}, \au{Caulfield, C.~P.}, \au{Dalziel, S.~B.} \& \au{Hunt,
  J. C.~R.}} \yr{2006}  \at{Time-dependent plumes and jets with decreasing
  source strengths}.  \jt{J. Fluid Mech.}  \bvol{563},  \pg{443--461}.

\bibitem[Schatzman(1978)]{Schatzman1978}
{\sc \au{Schatzman, M.}} \yr{1978}  \at{The integral equations for round
  buoyant jets in stratified flows}.  \jt{J. Appl. Math. Phys. (ZAMP)}
  \bvol{29},  \pg{608--630}.

\bibitem[Sillero {\em et~al.\/}(2013)Sillero, Jimenez \& Moser]{Sillero2013}
{\sc \au{Sillero, J.A.}, \au{Jimenez, J.} \& \au{Moser, R.D.}} \yr{2013}
  \at{One-point statistics for turbulent wall-bounded flows at {R}eynolds
  numbers up to $\delta^+\approx2000$}.  \jt{Phys. Fluids}  \bvol{25},
  \pg{105102}.

\bibitem[Silva {\em et~al.\/}(2018)Silva, Zecchetto \& da~Silva]{Silva2018}
{\sc \au{Silva, T.~S.}, \au{Zecchetto, M.} \& \au{da~Silva, C.~B.}} \yr{2018}
  \at{The scaling of the turbulent/non-turbulent interface at high {R}eynolds
  numbers}.  \jt{J. Fluid Mech.}  \bvol{843},  \pg{156--179}.

\bibitem[Sreenivasan {\em et~al.\/}(1989)Sreenivasan, Ramshankar \&
  Meneveau]{Sreenivasan1989}
{\sc \au{Sreenivasan, K.~R.}, \au{Ramshankar, R.} \& \au{Meneveau, C.}}
  \yr{1989}  \at{Mixing, entrainment and fractal dimensions of surfaces in
  turbulent flows}.  \jt{Proc. Roy. Soc. A -Math. Phys. Eng. Sci.}
  \bvol{421}~(1860),  \pg{79}.

\bibitem[Sullivan {\em et~al.\/}(1998)Sullivan, Moeng, Stevens, Lenschow \&
  Mayor]{Sullivan1998}
{\sc \au{Sullivan, P.~P.}, \au{Moeng, C.~H.}, \au{Stevens, B.}, \au{Lenschow,
  D.~H.} \& \au{Mayor, S.~D.}} \yr{1998}  \at{Structure of the entrainment zone
  capping the convective atmospheric boundary layer}.  \jt{J. Atmos. Sci.}
  \bvol{55},  \pg{3042--3064}.

\bibitem[Townsend(1976)]{Townsend1976}
{\sc \au{Townsend, A.~A.}} \yr{1976} {\em The structure of turbulent shear
  flow\/}.  \publ{Cambridge University Press.}

\bibitem[Turner(1962)]{TurJjfm1962a}
{\sc \au{Turner, J.~S.}} \yr{1962}  \at{The `starting plume' in neutral
  surroundings}.  \jt{J. Fluid Mech.}  \bvol{13}~(03),  \pg{356--368}.

\bibitem[Turner(1986)]{Turner1986}
{\sc \au{Turner, J.~S.}} \yr{1986}  \at{Turbulent entrainment: the development
  of the entrainment assumption, and its application to geophysical flows}.
  \jt{J. Fluid Mech.}  \bvol{173},  \pg{431--471}.

\bibitem[Watanabe {\em et~al.\/}(2016)Watanabe, Riley, De~Bruyn~Kops, Diamessis
  \& Zhou]{Watanabe2016}
{\sc \au{Watanabe, T.}, \au{Riley, J.}, \au{De~Bruyn~Kops, S.}, \au{Diamessis,
  P.} \& \au{Zhou, Q.}} \yr{2016}  \at{Turbulent/non-turbulent interfaces in
  wakes in stably stratified fluids.}  \jt{J. Fluid Mech.}  \bvol{797},
  \pg{R1}.

\bibitem[Watanabe {\em et~al.\/}(2018{\natexlab{{\em a\/}}})Watanabe, Riley,
  Nagata, Onishi \& Matsuda]{Watanabe2018}
{\sc \au{Watanabe, T.}, \au{Riley, J.}, \au{Nagata, K.}, \au{Onishi, R.} \&
  \au{Matsuda, K.}} \yr{2018{\natexlab{{\em a\/}}}}  \at{A localized turbulent
  mixing layer in a uniformly stratified environment}.  \jt{J. Fluid Mech.}
  \bvol{849},  \pg{245--276}.

\bibitem[Watanabe {\em et~al.\/}(2014)Watanabe, Sakai, Nagata, Ito \&
  Hayase]{Watanabe2014}
{\sc \au{Watanabe, T.}, \au{Sakai, Y.}, \au{Nagata, K.}, \au{Ito, Y.} \&
  \au{Hayase, T.}} \yr{2014}  \at{Enstrophy and passive scalar transport near
  the turbulent/non-turbulent interface in a turbulent planar jet flow.}
  \jt{Phys. Fluids}  \bvol{26}~(10),  \pg{105103}.

\bibitem[Watanabe {\em et~al.\/}(2018{\natexlab{{\em b\/}}})Watanabe, Zhang \&
  Nagata]{Watanabe2018a}
{\sc \au{Watanabe, T.}, \au{Zhang, X.} \& \au{Nagata, K.}}
  \yr{2018{\natexlab{{\em b\/}}}}  \at{Turbulent/non-turbulent interfaces
  detected in dns of incompressible turbulent boundary layers.}  \jt{Physics of
  Fluids}  \bvol{30}~(3),  \pg{035102}.

\bibitem[Wells {\em et~al.\/}(2010)Wells, Cenedese \& Caulfield]{Wells2010}
{\sc \au{Wells, M.}, \au{Cenedese, C.} \& \au{Caulfield, C.~P.}} \yr{2010}
  \at{The relationship between flux coefficient and entrainment ratio in
  density currents}.  \jt{J. Phys. Oceanogr.}  \bvol{40}~(12),
  \pg{2713--2727}.

\bibitem[Westerweel {\em et~al.\/}(2005)Westerweel, Fukushima, Pedersen \&
  Hunt]{Westerweel2005}
{\sc \au{Westerweel, J.}, \au{Fukushima, C.}, \au{Pedersen, J.~M.} \& \au{Hunt,
  J. C.~R.}} \yr{2005}  \at{Mechanics of the turbulent-nonturbulent interface
  of a jet}.  \jt{Phys. Rev. Lett.}  \bvol{95}~(17),  \pg{174501}.

\bibitem[Whitney(2005)]{WhiHboo1957a}
{\sc \au{Whitney, H.}} \yr{2005} {\em Geometric Integration Theory\/}.
  \publ{Dover}.

\bibitem[Woodhouse {\em et~al.\/}(2016)Woodhouse, Phillips \&
  Hogg]{WooMjfm2016a}
{\sc \au{Woodhouse, M.~J.}, \au{Phillips, J.~C.} \& \au{Hogg, A.~J.}} \yr{2016}
   \at{Unsteady turbulent buoyant plumes}.  \jt{J. Fluid Mech.}  \bvol{794},
  \pg{595--638}.

\bibitem[Woods(2010)]{Woods2010}
{\sc \au{Woods, A.~W.}} \yr{2010}  \at{Turbulent plumes in nature}.  \jt{Annu.
  Rev. Fluid Mech.}  \bvol{42},  \pg{391--412}.

\bibitem[Xu {\em et~al.\/}(1995)Xu, Fernando \& Boyer]{XuYphf1995a}
{\sc \au{Xu, Yunxiu}, \au{Fernando, Harindra J.~S.} \& \au{Boyer, Don~L.}}
  \yr{1995}  \at{Turbulent wakes of stratified flow past a cylinder}.
  \jt{Physics of Fluids}  \bvol{7}~(9),  \pg{2243--2255}.

\bibitem[Zhou \& Vassilicos(2017)]{Zhou2017}
{\sc \au{Zhou, Y.} \& \au{Vassilicos, J.C.}} \yr{2017}  \at{Related
  self-similar statistics of the turbulent/non-turbulent interface and the
  turbulence dissipation}.  \jt{J. Fluid Mech.}  \bvol{821},  \pg{440--457}.

\end{thebibliography}

\end{document}